\documentclass[]{aastex631}
\let\tablenum\relax
\usepackage{amsmath, multirow} 
\usepackage{mathrsfs}
\usepackage[scr=rsfs,cal=boondox]{mathalfa}

\usepackage{CJKutf8}
\usepackage{threeparttable}
\newcommand{\HeI}{He I}
\newcommand{\Ha}{H$\alpha$}
\newcommand{\Lya}{Ly$\alpha$}

\usepackage[many]{tcolorbox}

\begin{document}

\title{Helium escape in context: Comparative signatures of four close-in exoplanets}

\author[0009-0003-5882-9663]{Anna Ruth Taylor} 
\affiliation{Lunar and Planetary Laboratory, University of Arizona, Tucson, AZ 85721, USA}

\author[0000-0003-3071-8358]{Tommi T. Koskinen}
\affiliation{Lunar and Planetary Laboratory, University of Arizona, Tucson, AZ 85721, USA}

\author[0000-0001-9446-6853]{Chenliang Huang \begin{CJK*}{UTF8}{gbsn}(黄辰亮)\end{CJK*}}
\affiliation{Shanghai Astronomical Observatory, Chinese Academy of Sciences, Shanghai 200030, China}

\author[0000-0001-5097-4784]{Anthony Arfaux}
\affiliation{LIRA, Observatoire de Paris, Université PSL, CNRS, Sorbonne Université, Université Paris Cité, Meudon, France}

\author[0000-0002-5360-3660]{Panayotis Lavvas}
\affiliation{Laboratoire Enviromnements et Atmosphères Terrestres et Planétaires, Université de Reims Champagne Ardenne, Reims, France}
\affiliation{Institut d'Astrophysique de Paris, UMR CNRS 7095, Paris, France}

\begin{abstract}
Observations of escaping atmospheres on close-in exoplanets show a wide range in the strength and morphology of He I 10830~\AA\ and H I absorption. Scaling relations attempt to link the He I signal to XUV irradiation, mass loss, and bulk planetary parameters. We test these relations with a comparative analysis of HD209458b, HD189733b, HD149026b, and GJ1214b using a 1D hydrodynamic, multi-species, full-atmosphere escape model. For the benchmark HD209458b, our previously validated solution reproduces the observed He I and H$\alpha$ transit depths without imposing composition constraints. HD189733b exhibits comparable He I depths, but the broadest reported profiles require $\sim$12 km s$^{-1}$ of additional non-thermal broadening, whereas more recent measurements are narrower, consistent with our predictions. For HD149026b, despite similar system properties, our model shows that higher gravity suppresses escape and enhances diffusive separation, depleting helium at high altitudes and yielding extremely weak He I absorption. For the sub-Neptune GJ1214b, H/He-only models overestimate He I absorption; including H$_2$ and its ions (H$_2^+$, H$_3^+$, HeH$^+$) lowers the escape rate and modifies the ion/electron balance, reducing the metastable helium densities. Compared against scaling relations, HD189733b observations and our HD149026b prediction fall below the trend, whereas some observations of HD209458b and GJ1214b are consistent; however, the observed transit depths are variable. Across all targets, we find diffusive separation of helium and hydrogen, which may explain why sub-solar He/H ratios are often required in simplified models. We conclude that interpreting He I and H$\alpha$ absorption requires first-principles models that include self-consistent temperature and velocity profiles, multi-species transport, and molecular chemistry.
\end{abstract}


\section{Introduction} \label{sec:intro}
\noindent
Exoplanets orbiting close to their host stars receive intense stellar X-ray and ultraviolet (XUV) radiation that drives atmospheric escape. This process sculpts the observed exoplanet population by altering planetary radii, compositions, and atmospheric retention over time. Features such as the radius valley near 1.8 R$_\oplus$ \citep{Fulton2017AJ....154..109F} and the hot Neptune desert \citep{Mazeh2016A&A...589A..75M} can be explained by models of photoevaporation and Roche lobe overflow \citep[e.g.,][]{Owen2017ApJ...847...29O, 2022ApJ...929...52K}, highlighting the importance of understanding atmospheric escape in shaping planet demographics. Hot Jupiters and warm Neptunes provide particularly valuable laboratories for escape studies because their extended atmospheres can be probed observationally. Early detections of upper atmosphere escape relied on far-ultraviolet transit observations with the Hubble Space Telescope, primarily targeting H Ly$\alpha$ \citep[e.g.,][]{v-m2003Natur.422..143V, Ben-Jaffel2010ApJ...709.1284B}. However, Ly$\alpha$ is strongly absorbed by the interstellar medium, cannot be observed from the ground, and is strongly influenced by stellar activity, limiting both the number and precision of detections. More recently, ground-based and space-based detections of He I 10830~\AA\ and H I Balmer absorption have emerged as powerful complementary probes \citep[e.g.,][]{Spake2018Natur.557...68S, Casasayas2020A&A...640C...6C, Krishnamurthy_2024}. These lines trace the dominant escaping species, H and He, across altitudes spanning the middle atmosphere to the exosphere, while avoiding strong ISM attenuation and strong stellar contamination effects compared to other probes like  \Lya  ~\citep{Cauley2018AJ....156..189C}, and can be observed from the ground \citep{Indriolo2009ApJ...703.2131I}. Quantifying He I and H$\alpha$ absorption directly constrains the thermal structure, ionization balance, and escape efficiencies that population models require to reproduce features such as the radius valley and the hot Neptune desert. In practice, line depths and profiles translate into mass-loss rates and heating efficiencies, anchoring the timescales over which envelopes are removed and thereby setting the observed distribution of radii and compositions \citep[e.g.,][]{Owen2017ApJ...847...29O, 2022ApJ...929...52K}. Because the efficiency and modality of escape controls whether close-in sub-Neptunes retain H/He or are stripped to high-$\mu$ atmospheres, robust interpretation of \HeI\ and \Ha\ directly informs population features such as the radius valley and hot-Neptune desert by connecting present-day line depths to long-term mass-loss histories.

Population-level studies suggest that He I transit depths may correlate with the incident stellar XUV flux that ionizes He (5-504 \AA) \citep{Sanz-Forcada2025A&A...693A.285S} or mass-loss rate \citep{ Ballabio2025MNRAS.537.1305B}. However, substantial scatter—and even null trends with global parameters such as age, baseline transit depth, or $J$ magnitude \citep{Linssen2024arXiv241003228L, Allan2025arXiv250402578A}—indicates that physics missing from simple parameterizations (composition, multi-species transport, detailed energy balance, and 3D circulation) is critical. Observationally, He I absorption can probe compact thermospheres \citep{Alonso-Floriano2019} or extended tails \citep{Spake2018Natur.557...68S, Wang2021ApJ...914...98W, Zhang_Knutson_Wang_Dai_Barragan_2022}, underscoring system-to-system diversity. For the benchmark case of HD~209458b, earlier fits invoked subsolar He/H or ad-hoc diffusive separation to explain modest He I depths \citep{Lampon2020A&A...636A..13L, Biassoni2024A&A...682A.115B}. In contrast, recent self-consistent modeling \citep{Xing2023ApJ...953..166X,Taylor2025ApJ...989...68T} showed that modest diffusive separation that emerges naturally, together with a realistic thermal structure set by energy balance and a $\mu$bar-level lower boundary—reproduces both He I and H$\alpha$ transit depths without the need for a subsolar helium abundance in the planet's envelope. \citet{Taylor2025ApJ...989...68T}  also demonstrated how the degeneracy between temperature, velocity structure, and mass loss rate in frequently used Parker wind models \citep[e.g.,][]{Dossantos2022A&A...659A..62D, Alonso-Floriano2019, oklop2018ApJ...855L..11O} can lead to incomplete interpretations of observations due to the omission of key processes (e.g., multi-species transport and coupling to the lower atmosphere). This new model also included revised reaction rate coefficients for processes controlling the metastable helium population that impact the He I transit depths. Motivated by these results, we adopt the 1D multi-species model from \cite{Taylor2025ApJ...989...68T} to study different exoplanet systems and identify general drivers of the transit depths.

We focus on four representative planets—HD~209458b, HD~189733b, HD~149026b, and GJ~1214b—chosen to span distinct host-star environments (G, K, and M types) and escape regimes. Under a proposed scaling law of the form
\begin{equation}
\mathrm{EW}\,R_\star^2 \;\propto\; \frac{F_{\rm XUV, He}\,R_p^{2}}{\Phi_p},
\end{equation} where EW is the equivalent width of the He I 10830 \AA\ line, $R_\star$ is the stellar radius, $F_{\rm XUV, He}$ is the incident stellar XUV flux at wavelengths of 1-504 \AA\ that ionizes He, $R_p$ is the planet radius, and $\Phi_p$ is the gravitational potential of the planet \citep{Sanz-Forcada2025A&A...693A.285S}. The three hot Jupiters (HD~209458b, HD~189733b, HD~149026b) are expected to produce He I 10830~\AA\ EWs of $\sim$ 5.3 m\AA , 25 m\AA , and 1.9 m\AA, respectively, yet observations differ from these predictions. HD~209458b orbits a relatively quiet G0\,V star and exhibits a moderate but variable He I transit depth; HD~189733b orbits an active K2\,V star with enhanced XUV irradiation and reported He I 10830~\AA\ transit EWs for this planet span 1.88–12.8 m\AA\ \citep{Salz_2018, Guilluy_2020, Allart_2023, masson2024probing, Orell-Miquel_2025}. Additionally, the EW of 12.8 m\AA\ is broader than expected from thermal and radial escape velocity broadening \citep{Salz_2018, Guilluy_2020, Allart_2023, masson2024probing}.  Finally, HD~149026b, a hot Jupiter around a G0\,IV subgiant, yields only non-detections to date \citep{Biassoni2024A&A...682A.115B}. These contrasts motivate a process-level explanation beyond simple scaling relations. Host-star spectral energy distributions (SEDs) differ across G/K types in both XUV and NUV, which set the production (He I ground-state photoionization $\rightarrow$ He$^+$) and loss (photoionization of the 2$^3$S state) channels for the excited He triplet; coupled with gravity and transport, these SED differences can modulate the observed He I signal.

To explore a regime distinct from hot Jupiters, we also include the warm sub-Neptune GJ~1214b orbiting an M4.5 dwarf. M-dwarf SEDs generally deliver low NUV but variable and uncertain FUV/EUV fluxes from the transition region and the corona at the planet, altering both the ionization balance and the metastable helium budget. Although H/He-only models have sometimes been applied to sub-Neptunes \citep{Lampón_2021, Orell_2022}, past escape studies for Uranus-like atmospheres show that, even in the case of an envelope dominated by hydrogen and helium, H$_2$ can persist to high altitudes with very different thermal and chemical structures from hot Jupiters \citep{2022ApJ...929...52K}. Additionally, GJ 1214 b’s muted transmission spectrum points to a higher mean molecular weight in the atmosphere than generally for HJs \citep{Lavvas2019ApJ...878..118L, Lavvas_Paraskevaidou_Arfaux_2024}. GJ1214b is also interesting because the observed He I transit depths are variable \citep{Kasper_2022, Orell_2022, Allart_2023, masson2024probing} and the scaling relation predicts an EW of $\sim$50 m\AA. Modeling these four systems within a unified, self-consistent framework allows us to disentangle the roles of host-star SED, gravity, diffusion, and composition in shaping the helium and hydrogen diagnostics.


\section{Methods}\label{sec:methods}

\noindent
Here, we summarize our methods to simulate the atmospheres and calculate transit depths for these planets. The stellar and planetary parameters we use for each system are given in Table \ref{tab:params}. Section \ref{sec:lowmid} describes the lower/middle atmosphere models that we use to set the lower boundary conditions for the upper atmosphere models. In Section \ref{sec:uppermodel}, we describe the multi-species model of hydrodynamic escape that we use to model the upper atmospheres. Section \ref{sec:stellar} describes the inputs we use to model stellar fluxes of each star. The XUV fluxes (1-912\AA) in Table \ref{tab:params} come directly from these SEDs and are the flux at the planet, as described in Section \ref{sec:stellar}. In Section \ref{sec:scat}, we describe the detailed balance model used to compute the excited state H populations. Finally, In Section \ref{sec:radtran}, we describe the methods that we use to simulate transit depths based on the output of the lower/middle atmosphere models and the upper atmosphere models. For more details on this modeling scheme, see \cite{Taylor2025ApJ...989...68T}.

\begin{deluxetable*}{lcccc}[!h]
\centering
\tablecaption{Planetary and Stellar Parameters \label{tab:params}}
\tablewidth{0pt}
\tablehead{
\colhead{Parameter} & 
\colhead{HD 189733b} & 
\colhead{HD 209458b} & 
\colhead{HD 149026b} & 
\colhead{GJ 1214b}
}
\startdata
Stellar Mass ($M_*$) & 0.840 $M_\odot$ & 1.148 $M_\odot$ & 1.271 $M_\odot$ & 0.178 $M_\odot$ \\
Stellar Radius ($R_*$) &  0.805 $R_\odot$ & 1.162 $R_\odot$ & 1.290 $R_\odot$ & 0.215 $R_\odot$ \\
Planet Mass ($M_p$) & 1.142 $M_J$ & 0.714 $M_J$ & 0.346 $M_J$ & 0.0257 $M_J$ \\
Planet Radius ($R_p$) & 1.138 $R_J$ & 1.380 $R_J$ & 0.610 $R_J$ & 0.245 $R_J$ \\
Semi-major Axis ($a$) & 0.03142 AU & 0.04747 AU & 0.04288 AU & 0.01490 AU \\
Spectra Type & K2V  & G0V  & G0IV  & M4.5 \\
F$_{\text{XUV, H}}$ (W m$^{-2}$) [1-912 \AA] & 30.02 & 1.61 & 1.01 & 0.483 \\
F$_{\text{XUV, He}}$ (W m$^{-2}$) [1-504 \AA] & 22.77 & 1.19 & 0.85 & 0.478 \\
\enddata
\tablecomments{System parameters for HD189733b are from \citep{Boyajian_2015},  HD 209458 and HD 149026b from \citet{Southworth2010MNRAS.408.1689S}, and for GJ 1214b from \citet{Cloutier_2021}.}
\end{deluxetable*}

\subsection{Lower and Middle Atmosphere} \label{sec:lowmid}

\noindent
Ensuring consistency with the properties of the lower and middle atmosphere can be important for accurately modeling the upper atmosphere and escape processes. It also provides the necessary link between the transit depths measured in the upper atmosphere and those at neighboring wavelengths. To this end, we use the lower and middle atmosphere model of \citet{10.1093/mnras/stab456}, which calculates the thermal and compositional structure from 1000 bar to $10^{-6}$ bar where the latter pressure defines the lower boundary of our upper atmosphere simulations. The model solves for photochemistry and temperature self-consistently, incorporating roughly 150 species (atoms, molecules, and ions) connected by more than 1600 reactions. Based on the predicted structure and composition, it also produces a theoretical transmission spectrum that includes Rayleigh scattering, molecular absorption, and Mie scattering by aerosols when present. The dominant opacity sources include \(\text{H}_2\), \(\text{H}_2\text{O}\), \(\text{CH}_4\), \(\text{NH}_3\), \(\text{HCN}\), \(\text{H}_2\text{S}\), \(\text{CO}_2\), Na, K, collision-induced absorption by \(\text{H}_2-\text{H}_2\) and \(\text{He}-\text{H}_2\), along with numerous atomic metal lines, with cross-sections following \citet{Lavvas2021MNRAS.502.5643L}. Further description of the model is given in \citet{Arfaux_2022, Arfaux2023MNRAS.522.2525A, Lavvas_Paraskevaidou_Arfaux_2024}, and the references therein.

The lower/middle atmosphere model provides the lower boundary conditions for the escape simulations, setting both the temperature and altitude at 1 $\mu$bar. The continuum in the vicinity of the He I 10830 \AA\ and H$\alpha$ lines generally probes pressures deeper than the thermosphere and is primarily determined by water absorption along with cloud and haze extinction. Because of this, the level population calculations must extend smoothly below the upper atmosphere boundary. For metastable helium, we assume a balance between production via recombination and losses through radiative decay and Penning ionization, constrained by our detailed balance calculations. We also examined additional excitation pathways, such as Penning ionization with H$_2$, but find that they contribute negligibly to the overall He I population in these deeper layers. For excited hydrogen in the lower and middle atmosphere, see Section \ref{sec:scat}.

\subsection{Upper Atmosphere Escape Model} \label{sec:uppermodel}

\noindent
We employ a one-dimensional model of the thermosphere–ionosphere system and hydrodynamic escape tailored for close-in exoplanets \citep{2022ApJ...929...52K,Taylor2025ApJ...989...68T}. The code integrates the coupled, time-dependent equations of mass, momentum, and energy conservation in the radial direction. Physical processes included are photoionization by stellar XUV radiation, thermal ionization, recombination, charge exchange, advection and escape of both ions and neutrals, molecular and atomic diffusion, viscous drag, adiabatic heating and cooling, thermal conduction, and viscous dissipation. Radiative cooling is included through recombination, H I line emission \citep{Huang__2023}, and H$_3^+$ cooling \citep[e.g.,][]{2022ApJ...929...52K}, as appropriate. Photochemistry and the population kinetics of helium are solved using the open-source Kinetic Preprocessor (KPP-3.0.0), which is fully integrated with the escape model \citep{KPP}. The principal tunable inputs in this work are the photoelectron heating efficiency, the eddy diffusion coefficient $K_{zz}$, the incident stellar activity/XUV level, and the adopted lower-boundary composition (H/He vs. inclusion of H$_2$/metals). For the eddy diffusion coefficient $K_{zz}$, we use a value consistent with the lower and middle atmosphere model that is based on a recent assessment of eddy diffusion in exoplanet atmospheres \citep{Arfaux2023MNRAS.522.2525A}. A more complete description of the numerical implementation is provided by \citet{2022ApJ...929...52K}.

We consider two classes of upper atmosphere models in this work. Two of our simulations focus on hydrogen and helium only, including the species H, H(n=2), He, H$^+$, He$^+$, He$^{2+}$, He($2^3$S), and electrons. The reactions for these species are listed in Table 2 of \citet{Taylor2025ApJ...989...68T}, with the exception of H(n=2), which is handled separately (see Section~\ref{sec:scat}). In Sections \ref{sec:189} and \ref{sec:1214}, we also present a model that includes molecular hydrogen (H$_2$) and the ions H$_2^+$, H$_3^+$ and HeH$^+$ in addition to the atomic and ionic species listed above. These molecular species can influence both the mean molecular weight and the radiative cooling of the upper atmosphere, thereby altering the thermal structure and escape rate. The chemical network and reaction rates for the H$_2$-based model follow the implementation described in \citet{2022ApJ...929...52K}. In addition, we explicitly include H$_2$ photodissociation with branching between neutral dissociation (H$_2$ + $h\nu$ $\rightarrow$ H + H) and dissociative ionization channels, using wavelength-dependent branching ratios for the neutral branch \citep{Backx1976, Dujardin1987, Chung1993}; thermal dissociation is also included. 

Note that the composition of our upper atmosphere models is simpler than the composition of the lower and middle atmosphere models. This creates a discontinuity in composition across the lower boundary pressure of our escape model. In particular, the composition in the lower and middle atmosphere of GJ1214b is substantially incompatible with our upper atmosphere model that does not include molecules such as H$_2$O and CO. The purpose of our work here is to show that the inclusion of H$_2$ and related chemistry alone in the upper atmosphere has a significant impact on the observed transit depths. A comprehensive study of molecular chemistry and its impact on different transit depths probing the upper atmosphere is an obvious direction for future work. For HD149026b, the atoms only composition in the upper atmosphere model also conflicts with the lower and middle atmosphere model because the abundance of H$_2$ at the lower boundary pressure of the upper atmosphere model is predicted to be relatively high. The upper atmosphere model for this planet effectively assumes that H$_2$ is abruptly dissociated at the 1 $\mu$bar level. We do not include H$_2$ chemistry in the HD149026b upper atmosphere model since the predicted transit depths are not detectable and adding H$_2$ would only serve to somewhat reduce them further. We include H$_2$ for HD189733b because its lower and middle atmosphere remains H$_2$-dominated at 1$\mu$bar, unlike HD209458b where atomic hydrogen already dominates at this level. 


\subsection{Stellar Fluxes}
\label{sec:stellar}
\noindent HD~209458 is a G0-type star that is relatively inactive, with a coronal temperature similar to that of the quiet Sun \citep{Czesla_2017}. The Sun’s quiescent X-ray luminosity spans $10^{27}$–$10^{28}$ erg s$^{-1}$ \citep{Judge2003ApJ...593..534J}, and HD~209458 lies at the lower end of this range with a measured value of $\sim 1.6 \times 10^{27}$ erg s$^{-1}$ \citep{Czesla_2017}. To reproduce these conditions, we adopt a solar minimum spectrum from 2008, obtained from LISIRD\footnote{\url{https://lasp.colorado.edu/lisird/}} and scaled to the radius of HD~209458, as the baseline input spectrum for our model. To test how stellar activity influences the predicted He I (2$^3$S) and H$\alpha$ transit depths, we also employ solar average and solar maximum spectra based on TIMED SEE data, again scaled to the stellar radius of HD~209458. For each activity case, we combine LISIRD/TIMED SEE spectra at wavelengths shorter than 185 nm with a PHOENIX \citep{Husser2013A&A...553A...6H} model spectrum (T$_\text{eff}$ = 6075 K, log($g$) = 4.38, \citealt{del_Burgo_Allende_Prieto_2016}) at longer wavelengths. The connection between these two spectra is smooth. To calculate the H I level populations (Section~\ref{sec:scat}), we additionally require the Ly$\alpha$ flux. For this, we adopt the solar Ly$\alpha$ line profiles from \citet{LEMAIRE2005384}, scaled consistently to match the stellar activity level used in each case. Note that these are the same stellar inputs used in \cite{Taylor2025ApJ...989...68T}.

HD 189733 is a K2V-type star with moderate magnetic activity, and its high-energy spectrum has been studied extensively as part of the MUSCLES Treasury Survey \citep{France2016ApJ...820...89F, Loyd2016ApJ...824..102L, Youngblood2016ApJ...824..101Y}. For our model, we adopt the stellar spectrum of $\epsilon$ Eri, another K2V star, as a proxy for HD 189733 \citep{Sanz-Forcada2011A&A...532A...6S}. To calculate the H I level populations via resonant scattering (see Section~\ref{sec:scat}), we adopt the reconstructed Ly$\alpha$ profile of $\epsilon$ Eri from \citet{Youngblood2016ApJ...824..101Y}, scaled consistently with the full spectrum. This setup captures the moderately active nature of HD 189733 and allows us to assess its impact on He I and H$\alpha$ line formation in the planet’s escaping atmosphere.

HD 149026 is a G0IV-type star, and we use its observed stellar spectrum from the MUSCLES Treasury Survey (version 2.1) \citep{France2016ApJ...820...89F, Loyd2016ApJ...824..102L, Youngblood2016ApJ...824..101Y} as the input for our model. The MUSCLES database provides a reconstructed spectral energy distribution ($F_\lambda$) from X-rays through the infrared based on multi-wavelength observations and empirical models. We use the full MUSCLES spectrum across all wavelengths without modification. The spectrum is broken down as follows: the 1213.55–1218 \AA\ Ly$\alpha$ profile is reconstructed \citep{Youngblood2016ApJ...824..101Y}, and the 100.5–1169.5 \AA\ EUV is inferred from Ly$\alpha$ via the \citet{Linsky2014ApJ...780...61L} scaling, while the 2274.54–3112.98 \AA\ and 1301.5–1563.0 AA\ segments are direct HST/STIS observations. A scaled quiet-Sun proxy is used where there are gaps in data, and $\lambda \ge$ 5689.52~\AA~a PHOENIX photosphere is used. The Ly$\alpha$ line profile used for H I resonant scattering calculations (see Section~\ref{sec:scat}) is taken directly from the MUSCLES data products for HD 149026. This self-consistent input spectrum allows us to model the planet's upper atmosphere without the need for stellar analogs or synthetic models.

For GJ 1214, we use a custom stellar spectrum based on the MUSCLES composite spectrum for GJ 1214 and a scaled version of the GJ 436 model spectrum from \citet{Lavvas2019ApJ...878..118L}. The X-ray portion (0.5–10 nm) is taken directly from the MUSCLES spectrum, which is based on XMM-Newton observations modeled using the APEC code \citep{Lalitha2014ApJ...790L..11L, Loyd2016ApJ...824..102L}. The EUV region (10–117 nm) is poorly constrained due to low S/N in Ly$\alpha$ observations and was replaced with the GJ 436 spectrum scaled downward using band-specific factors: 0.194 (10–40 nm), 0.0198 (40–91 nm), and 0.168 (91–117 nm) based on the empirical scaling relationship from \cite{Linsky2014ApJ...780...61L}. The Ly$\alpha$ core (121.4–121.7 nm) was scaled by a factor of 0.113 to reflect the difference in reconstructed Ly$\alpha$ flux between GJ 1214 and GJ 436. For the FUV range (122–305 nm), we scale down the GJ 436 \cite{Lavvas2019ApJ...878..118L} spectrum consistently with the Ly$\alpha$ scaling factor, which is 0.113 given by the scaling relations in \cite{Linsky2014ApJ...780...61L}. We use the unmodified MUSCLES spectrum for GJ124 for wavelengths longer than 305 nm.  

We note that for all the host stars, we compared our SEDs to the \citet{Sanz-Forcada2025A&A...693A.285S} SEDs by integrating the 1–912 Å photon flux at the planet for each star. We find that the integrated 1–912 Å photon flux in our SEDs is slightly lower than that of the \citet{Sanz-Forcada2025A&A...693A.285S} synthetic spectra by a factor of 1.18 - 1.66.

\subsection{H I Level Populations} \label{sec:scat}

\noindent
In addition to modeling the He I 10830~\AA\ line, we also simulate H$\alpha$ absorption using a non-LTE framework for the hydrogen level populations \citep{2017ApJ...851..150H, Huang__2023} for the lower/middle and upper atmopshere. Previous studies have shown that excitation of the H($n=2$) state is dominated by radiative processes, with the H($2p$) population in particular set by the local Ly$\alpha$ radiation field \citep{2017ApJ...851..150H, 2019ApJ...884L..43G}. To capture this, we apply the Ly$\alpha$ Monte Carlo resonant scattering radiative transfer model of \citet{2017ApJ...851..150H} to compute the mean Ly$\alpha$ intensity in a plane-parallel atmosphere.  Because this calculation is computationally expensive, it is not practical to couple the non-LTE hydrogen model directly to the hydrodynamic escape simulations. Instead, we compute the Ly$\alpha$ radiation field separately using the output from the upper atmosphere model, then update the H($n=2$) density accordingly. The hydrodynamic and radiative transfer calculations are repeated iteratively until the temperature and electron density profiles converge, which typically requires three iterations. This approach ensures that the hydrogen level populations are consistent with the thermospheric conditions. We also note that ionization of H($n=2$) is a non-negligible source of free electrons, contributing to the thermal balance of the upper atmosphere. We follow the methodology of \citet{Huang__2023}; see that work for full derivations and implementation details.

\subsection{Simulated Transit Depths} \label{sec:radtran}

\noindent
To compare our atmospheric models with observations of H I and He I 10830~\AA\ absorption, we compute transit depths using the modeled pressure, temperature, and density profiles from both the upper and lower/middle atmosphere models. We calculate the transit depths by ray-tracing parallel chords through the modeled pressure–temperature–density structure, integrating the wavelength-dependent opacity along each impact parameter, and averaging the resulting transmissions across the stellar disk, following equation (11) of \citet{Brown2001ApJ...553.1006B}. More details on this calculation can be found in \cite{Taylor2025ApJ...989...68T}.
Line parameters for atomic transitions are taken from the NIST database\footnote{\url{https://physics.nist.gov/PhysRefData/ASD/lines_form.html}}. Here, we restrict our attention to the He I 10830~\AA\ triplet and the H$\alpha$ Balmer line; the lower/middle atmosphere continuum spectrum and associated molecular opacity sources are described in Section~\ref{sec:lowmid}.  

For the atomic line profiles, we calculate transit depths using a Voigt profile that includes both natural and thermal broadening. Additional broadening is incorporated by adding the line-of-sight outflow velocity to the Gaussian component, along with rotational broadening set by the planet’s spin. This simplified approach allows straightforward comparison with prior studies and provides clear constraints on the velocity broadening required to reproduce observations. We emphasize that transit depths may be influenced by three-dimensional dynamics, magnetic confinement, or stellar wind interactions \citep{Owen2020SSRv..216..129O}, none of which are included in our 1D framework. Nevertheless, 1D models remain valuable for exploring the physics of escape and identifying key inputs for more complex simulations. They also enable efficient surveys of population-level trends. Finally, we note that recent work suggests transmission spectra are often dominated by dayside-like conditions rather than pure terminator averages \citep{Jaziri2024A&A...684A..25J}. This is particularly relevant for hot, extended thermospheres where stellar radiation penetrates beyond the terminator onto the nightside, implying that a significant fraction of the escaping atmosphere exhibits dayside-like properties, even in the absence of redistribution by circulation \citep{Koskinen2007}.

\section{Results} \label{sec:results}
\noindent

\begin{figure}[h!]
    \centering
    \includegraphics[width=0.47\textwidth]{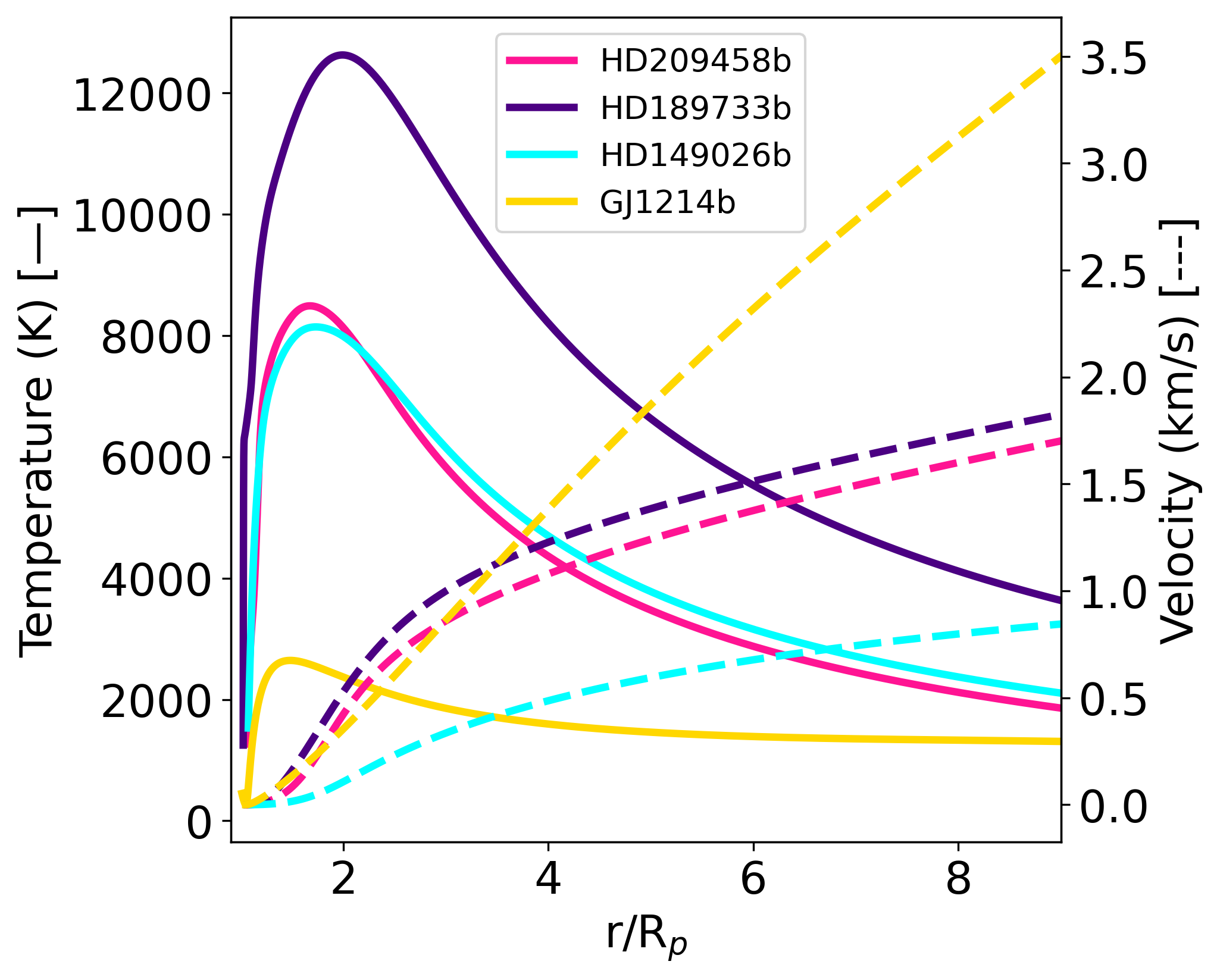}
    \includegraphics[width=0.4\textwidth]{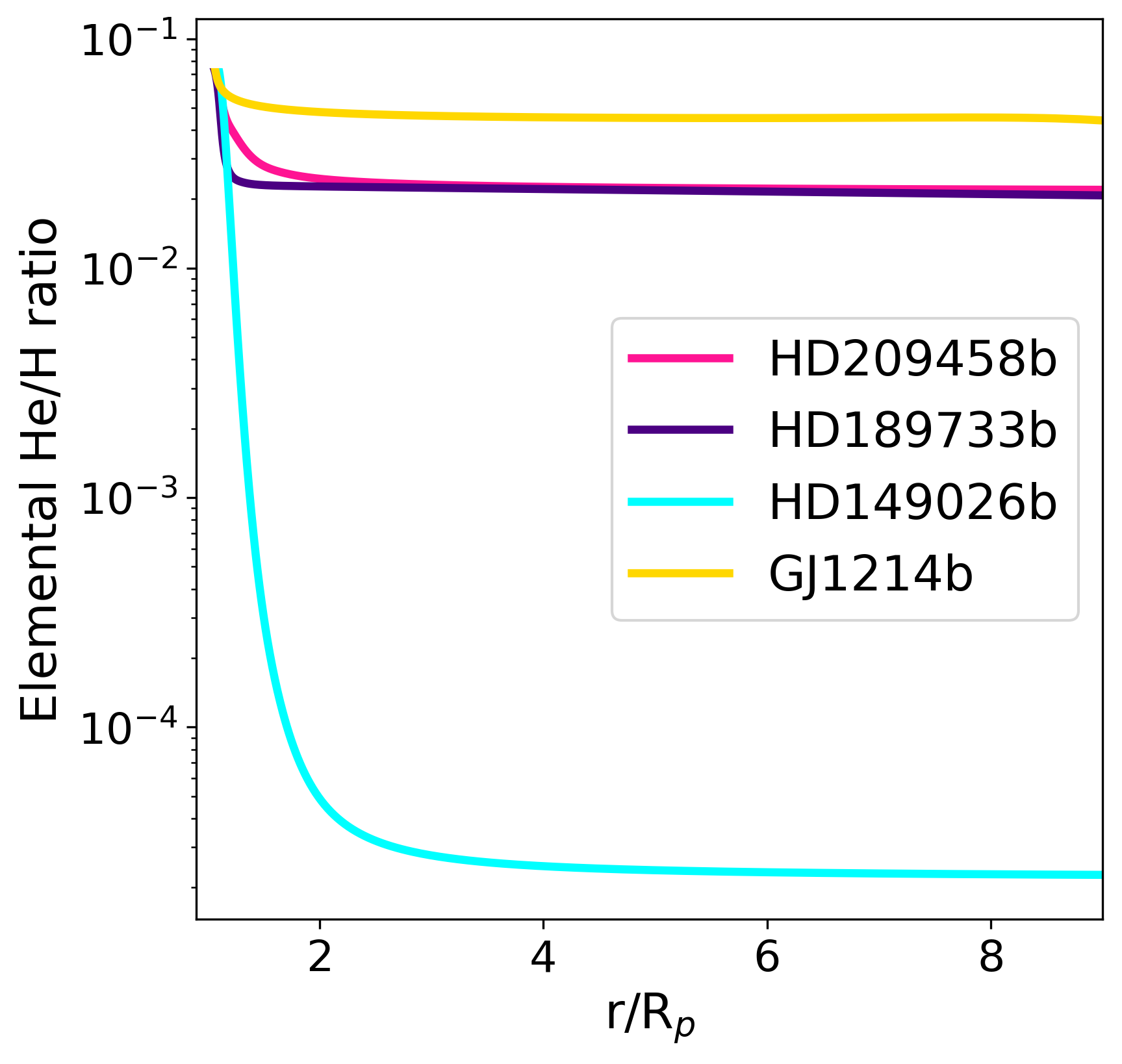}

    \caption{Upper atmosphere model results for our 4 planet sample. \textit{Left:} Temperature (solid lines) and outflow velocity profiles (dashed lines). \textit{Right:} Elemental He/H ratio as a function of radius.}
    \label{fig:all_tvcomp}
\end{figure}

\begin{figure}[h!]
    \centering
    \includegraphics[width=0.7\textwidth]{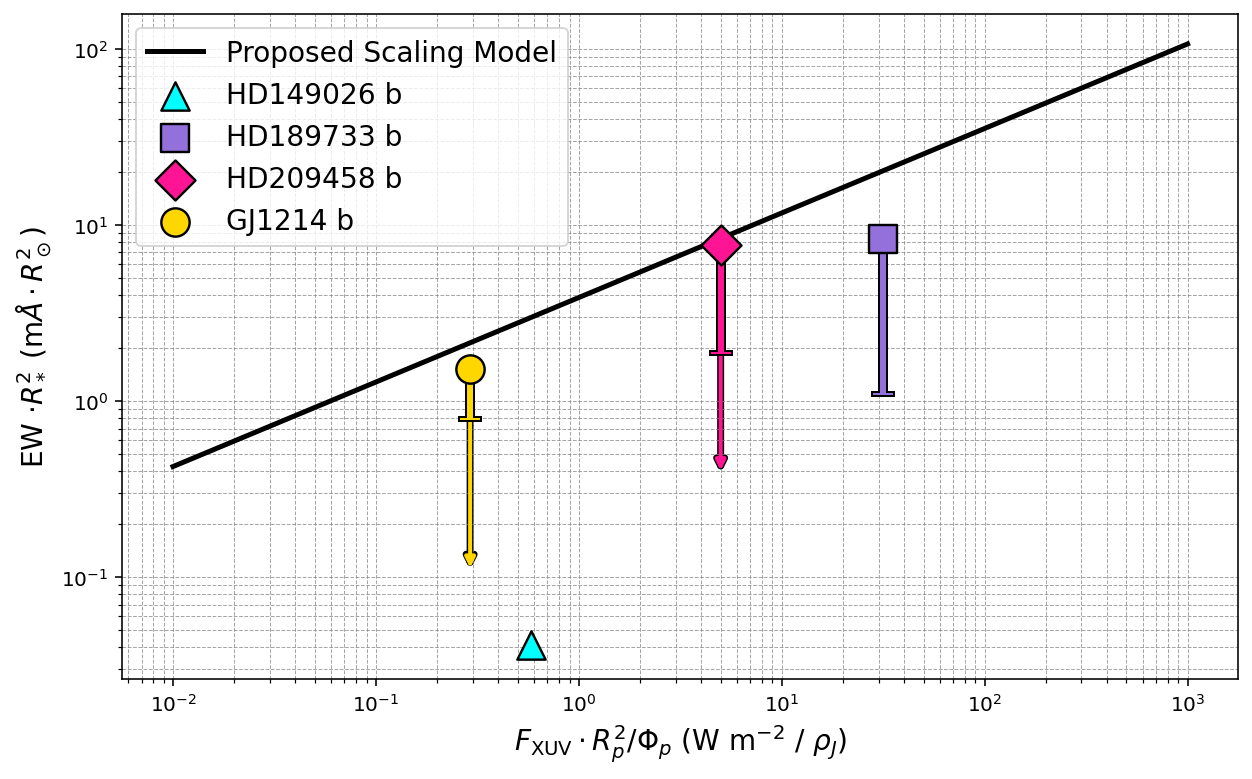}
    \caption{Recreation of the proposed scaling relation from \citet{Sanz-Forcada2025A&A...693A.285S}. Symbols indicate results for the four planets in this study. The points for HD209458b, HD189733b, and GJ1214b are based on observations, while the point for HD149026b is based on our model prediction. Lower error bar limits for GJ1214b and HD209458b are given by upper limits in \citet{masson2024probing}, and the lower limit for HD189733b is given by \citet{Orell-Miquel_2025}. The solid black line shows the predicted scaling of EW with $F_{\mathrm{XUV, He}}$$ \cdot R_p^2 / \Phi_p$.  Note that for the x-axis we use the $F_{\mathrm{XUV, He}}$ flux given in the last row of Table \ref{tab:params}}.
    \label{fig:scaling}
\end{figure}

Here, we present results from our atmospheric models for the planets, focusing on critical parameters such as temperature and velocity distributions, atmospheric composition, metastable helium production and loss processes, and transport mechanisms. Figure \ref{fig:all_tvcomp} shows the temperature and velocity (left), and the elemental He/H ratios (right) for our four model atmospheres. In order to compare our predictions with a recent scaling relation, Figure \ref{fig:scaling} shows where our predicted transit depths fall based on the scaling relation of \citet{Sanz-Forcada2025A&A...693A.285S}. For HD209458b, HD189733b, and GJ1214b, our modeled He I (and H$\alpha$, where applicable) transit depths fall within the ranges reported in the literature [HD209458b: He I and H$\alpha$ \citep[e.g.,][]{Alonso-Floriano2019, Winn2004, Astudillo-Defru2013A&A...557A..56A, Casasayas-Barris2021A&A...647A..26C}; HD189733b: He I \citep[e.g.,][]{Allart_2023, masson2024probing, Zhang_Knutson_Wang_Dai_Barragan_2022}; GJ1214b: mixed He I detections and non-detections \citep[e.g.,][]{Kasper_2022, Orell_2022, Allart_2023, masson2024probing}], whereas HD~149026b currently has no confirmed detection and only upper limits; our model thus serves as a prediction consistent with present constraints \citep{Biassoni2024A&A...682A.115B}. Because measured depths vary across epochs and instruments, Figure \ref{fig:scaling} also overlays the reported observational ranges (or upper limits) for each planet.  We return to this figure in the following subsections to highlight each planet’s departure or agreement with this scaling relation. 

\subsection{HD209458b} \label{sec:209}

\begin{figure}[h!]
    \centering
    \includegraphics[width=0.39\textwidth]{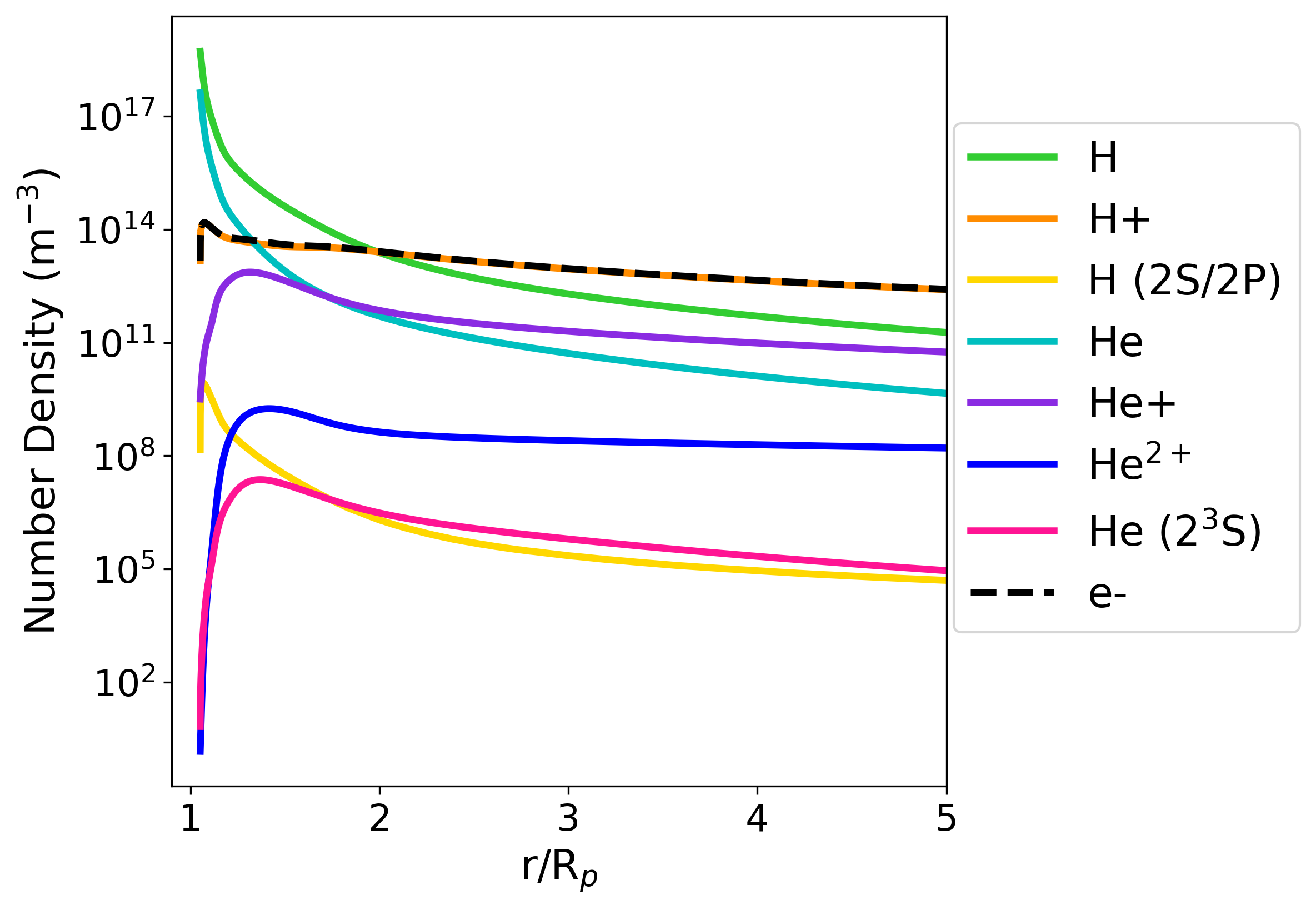}
    \includegraphics[width=0.29\textwidth]{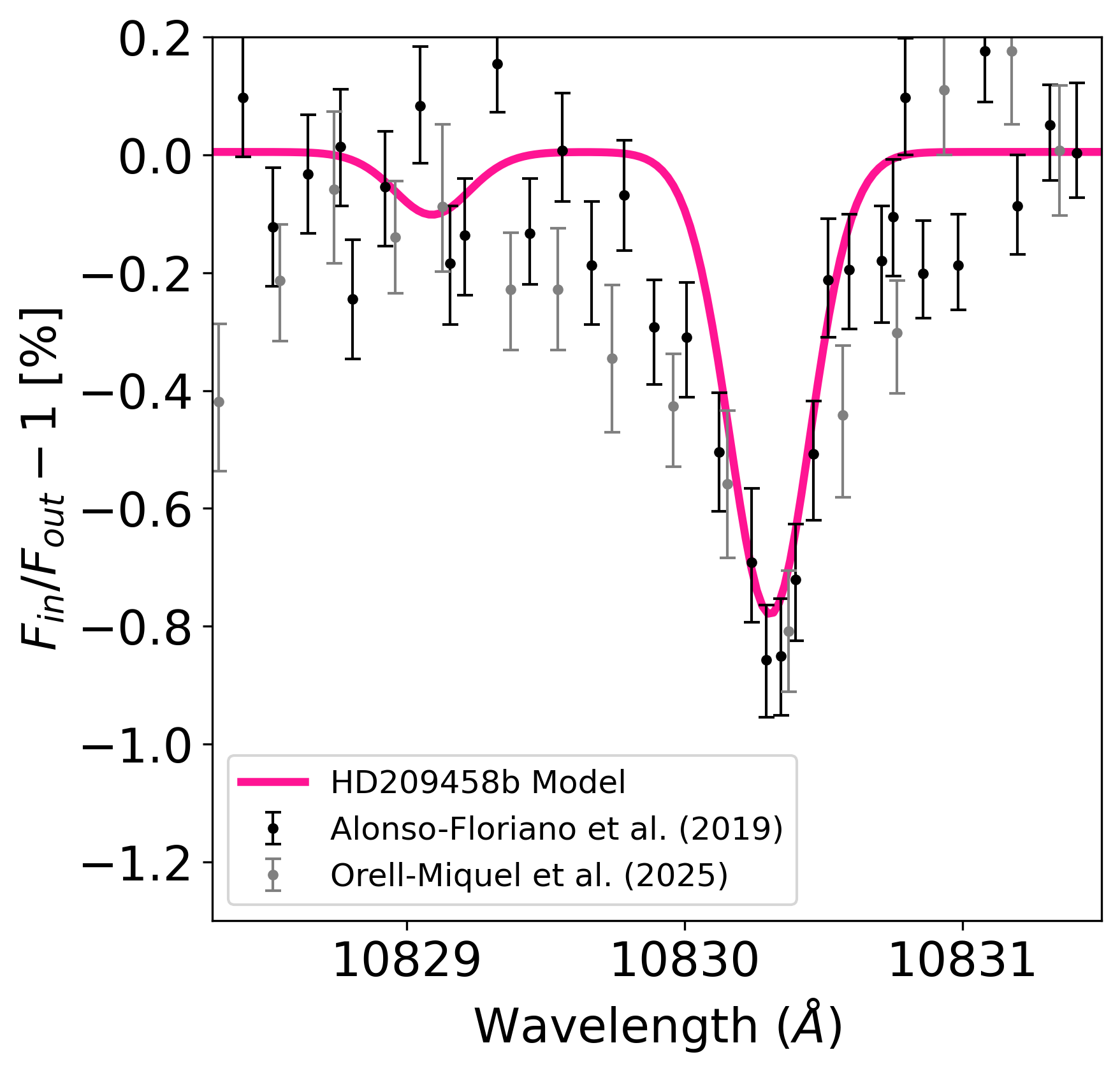}
    \includegraphics[width=0.29\textwidth]{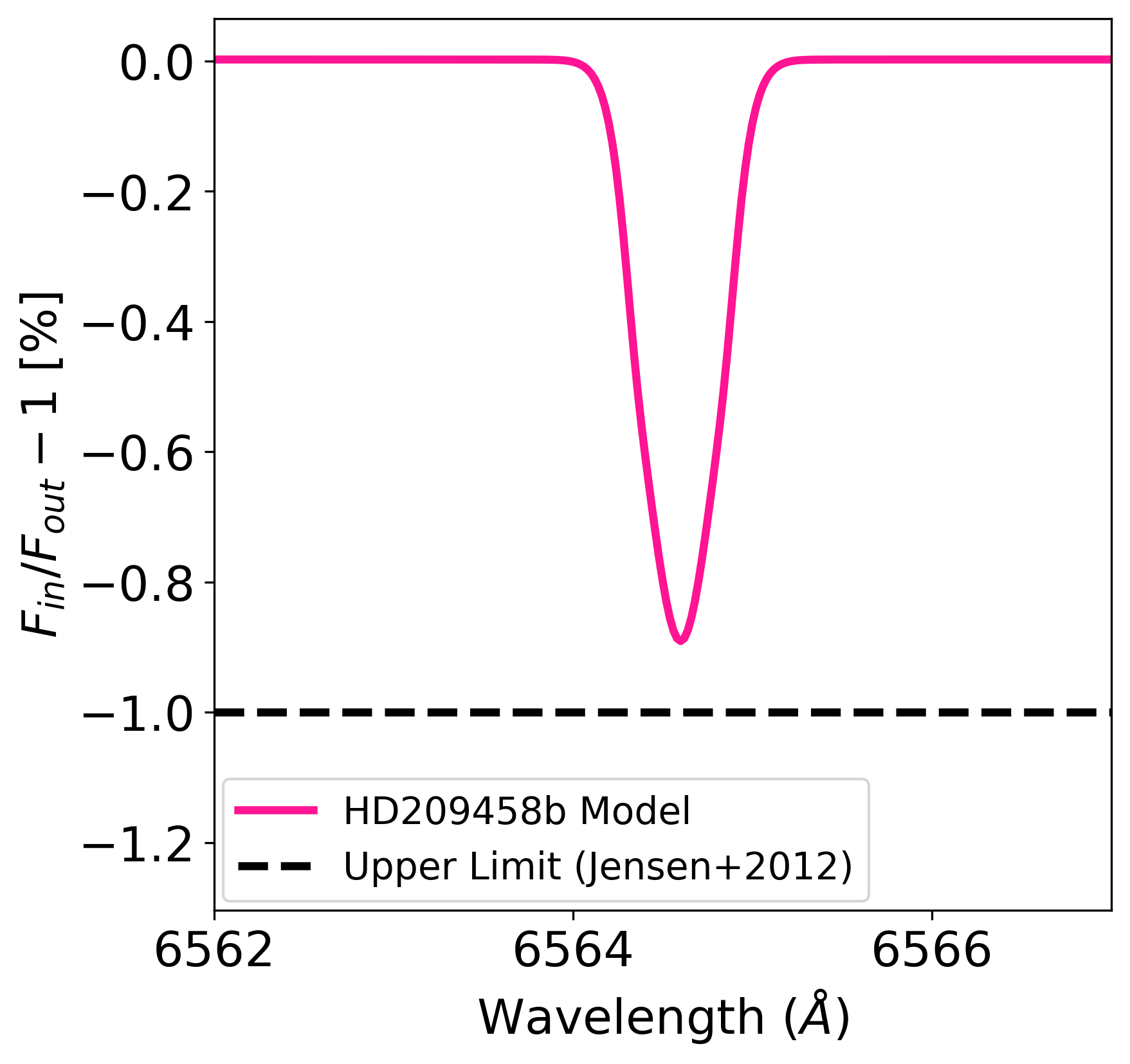}
    \caption{\textit{Left:} Composition results for HD 209458b in the upper atmosphere. Number densities of H, H$^+$, He, He$^+$, He I (2$^3$S), and electrons. \textit{Middle:}  Modeled He I transit depth spectra for HD 209458b compared with the observations from \cite{Alonso-Floriano2019} and \citep{Orell-Miquel_2025}.  \textit{Right:} H$\alpha$ Balmer line compared with the upper limit from \cite{jenesen2012ApJ...751...86J}. The model results are the same as those in \citet{Taylor2025ApJ...989...68T} for ease of comparison with the other planets modeled in this work.}
    \label{fig:hd209_transit}
\end{figure}

HD~209458b serves as the benchmark system for our modeling framework. We modeled HD 209458b in detail in \citet{Taylor2025ApJ...989...68T}, using our self-consistent full atmosphere multi-species model. Our best-fit H/He only model matches the observations with a photoelectron heating efficiency of 40\% and yields a mass-loss rate of $\dot{M} = 1.9 \times 10^{10}$ g/s. The temperature profile peaks near 8200 K at 1.7 $R_p$, and the outflow velocity is 1.1 km/s at 4 R$_p$. Note that we do not include Roche lobe overflow in this model, since it is not necessarily a good assumption for a global average. The pink lines in Figure~\ref{fig:all_tvcomp} shows the thermal structure and He/H ratio of the atmosphere. The He/H ratio shows diffusive separation, declining from 8\% at the base to 2.5\% at high altitudes. As shown in Figure~\ref{fig:hd209_transit}, the modeled He~I~10830 \AA\ transit depth reaches $\sim$0.8\%, closely matching the observed line from \cite{Alonso-Floriano2019} and \cite{Orell-Miquel_2025}. The H$\alpha$ absorption depth reaches $\sim$0.9\%, consistent with the upper limit reported in \citet{Winn2004} and \citet{jenesen2012ApJ...751...86J}. Other H$\alpha$ constraints in the literature are not fully consistent with our best-fit model \citep{Astudillo-Defru2013A&A...557A..56A, Casasayas-Barris2021A&A...647A..26C}. See \cite{Taylor2025ApJ...989...68T} for an in-depth discussion of these discrepancies. 
The H$\alpha$ observations for HD209458b are not contemporaneous with He I observations and could exhibit variability, including effects due to stellar variability. Coordinated, high-resolution, simultaneous campaigns will be needed to refine the H$\alpha$ constraints and probe variability.

    


\subsection{HD189733b} \label{sec:189}

\begin{figure}[h!]
    \centering
    \includegraphics[width=0.38\textwidth]{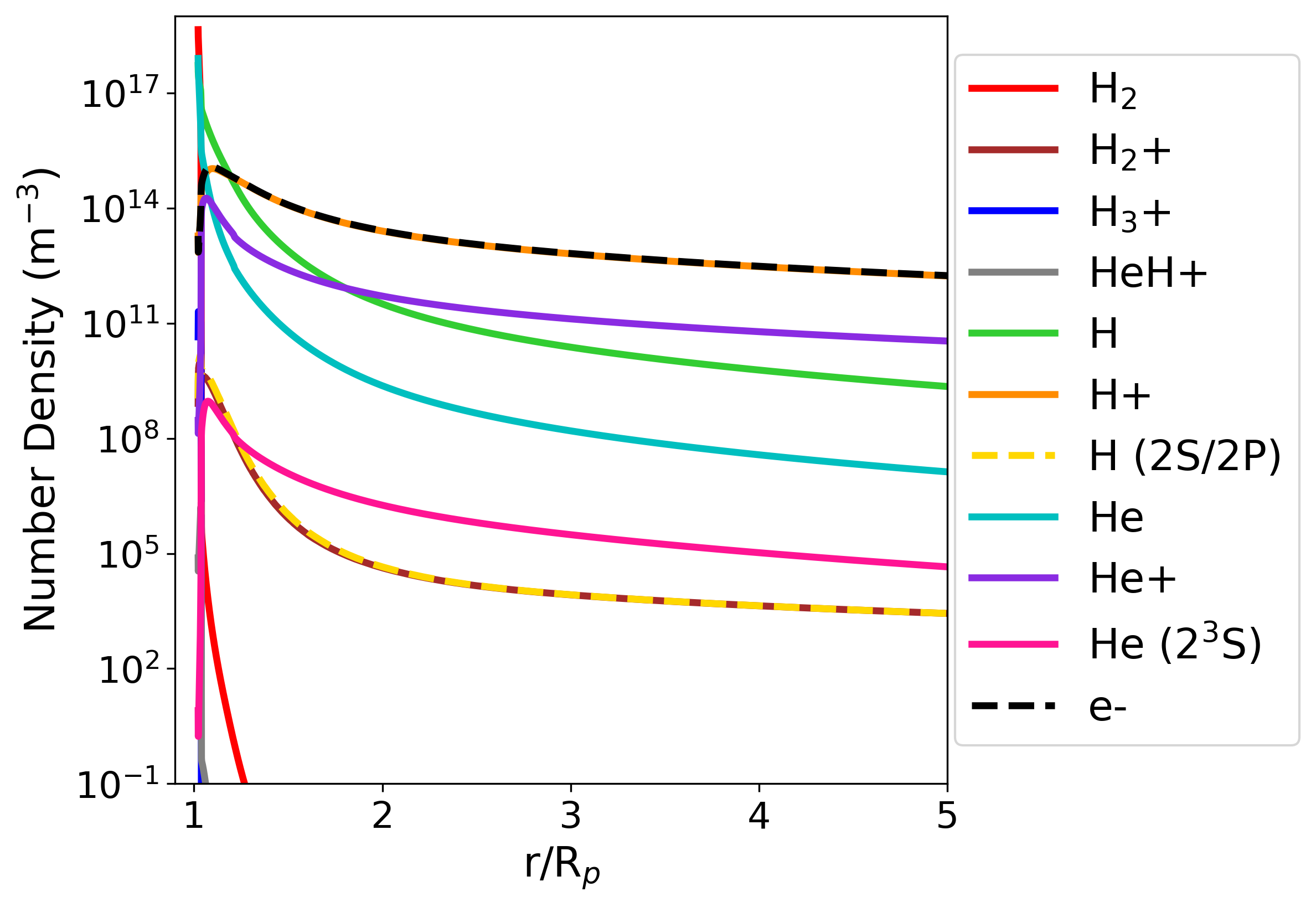}
    \includegraphics[width=0.28\textwidth]{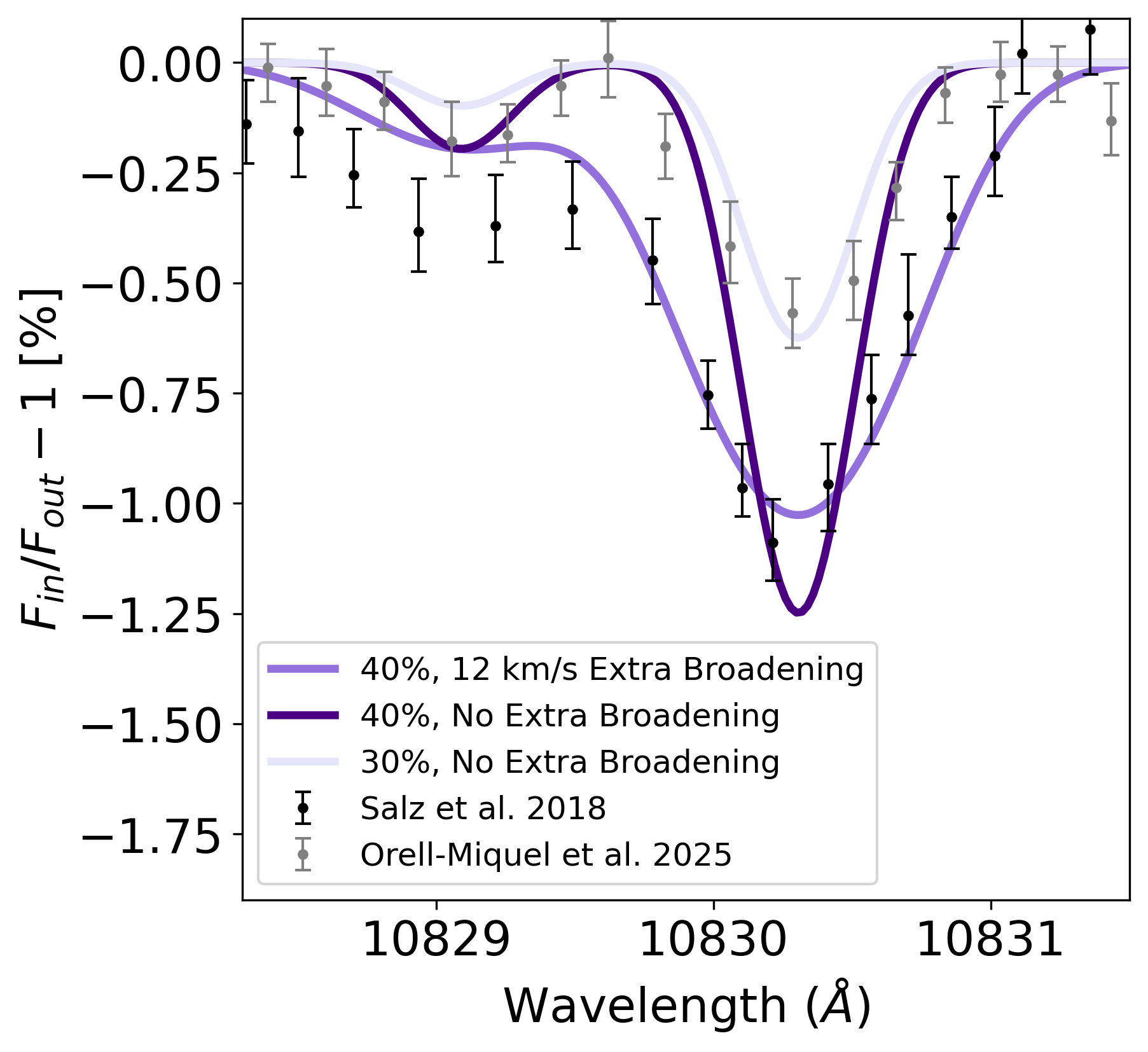}
    \includegraphics[width=0.30\textwidth]{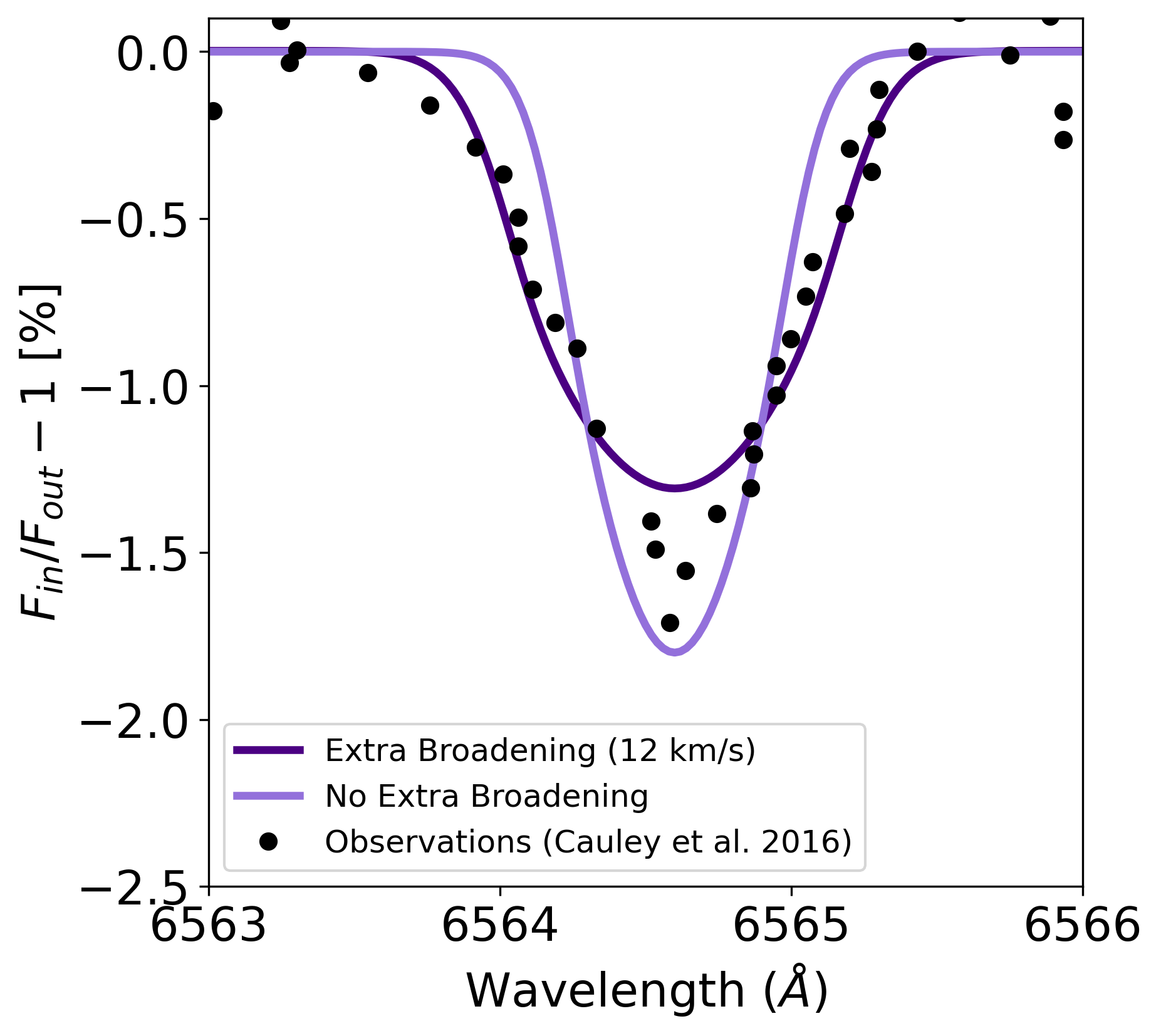}
    \caption{Composition results for HD 189733b in the upper atmosphere. \textit{Left:} Number densities of major species including H, H$^+$, He, He$^+$, He I (2$^3$S), H$_2$,  H$_2$+, H$_3$+, and electrons. \textit{Middle:} Model He I 10830 \AA transit depth compared with observations, including a model with 40\% photoelectron heating efficiency and 12 km/s velocity broadening (dark purple), consistent with the observations from \cite{Salz_2018}, a model with no extra broadening (purple), and a model with 30\% photoelectron heating efficiency and no extra broadening, consistent with observations from \cite{Orell-Miquel_2025} (light purple). \textit{Right:} H I Balmer $\alpha$  line including 12 km/s velocity broadening (dark purple) and no extra broadening (purple). Both features are consistent with observations from  and \cite{Cauley_Redfield_Jensen_Barman_2016}}
    \label{fig:hd189_transit}
\end{figure}

HD 189733b is a well-studied hot Jupiter orbiting an active K-dwarf, and one of the best systems for joint analysis of H I and He I upper atmosphere absorption, because of its proximity to our solar system. The left panel with purple lines of Figure~\ref{fig:all_tvcomp} shows the temperature and composition profiles in our best-fit model. In order to match the observations, we use a photoelectron heating efficiency of 40\% that leads to a mass-loss rate of $\dot{M} = 7.1 \times 10^{9}$ g/s. The temperature profile peaks at 13,000 K near 1.9 $R_p$. As shown in the right panel of Figure~\ref{fig:all_tvcomp}, HD 189733b exhibits diffusive separation of helium and hydrogen, with the He/H ratio decreasing from 8\% at the base to 2\% at high altitudes. The left panel of Figure \ref{fig:hd189_transit} shows the composition for our model of HD189733b. We included H$_2$ in this model for consistency with the lower/middle atmosphere, but H$_2$ is quickly dissociated in the upper atmosphere model and therefore has a small effect on the transit depths or mass-loss rate. 


The observed He I 10830~\AA\ line core depth of HD~189733b is comparable to that of HD~209458b (e.g., \citealt{Alonso-Floriano2019, Salz_2018, Cauley_2017}), but the profile can be markedly broader. The middle panel of Figure~\ref{fig:hd189_transit} shows various representative published line profiles for HD~189733b. Several prior analyses have matched the HD~189733b width by introducing ad-hoc (micro)turbulent or Gaussian broadening terms; here we show that this step remains necessary even in our self-consistent framework. Using the same modeling setup as for HD~209458b—including coupling to the lower/middle atmosphere, energy-balance thermal structure, multi-species diffusion, and non-LTE H($n{=}2$) populations—and adopting a comparable photoelectron heating efficiency to HD 209458b, our HD~189733b model reproduces the line core depth without invoking subsolar He/H or other composition priors. Our simulated transit depth accounts for thermal and natural broadening, line-of-sight outflow, and rotational broadening. Despite these ingredients and the known system differences from HD 209458b (K2\,V host with enhanced XUV at $a{=}0.031$~AU, higher gravity and different $R_p/R_\star$), the modeled profile remains too narrow. Matching the observed width from \cite{Salz_2018}, that are consistent with many other observations \citep[i.e.,][]{Allart_2023, masson2024probing, Zhang_Knutson_Wang_Dai_Barragan_2022}, requires additional non-thermal velocity dispersion of $\sim$12~km\,s$^{-1}$, significantly larger than the thermal, bulk-flow, or rotational broadening predicted by the model. We discuss the possible physical origin of this broadening in Section \ref{sec:dis}. Note, however, that recent observations from \citet{Orell-Miquel_2025} show a much shallower, narrower transit depth that we match with a 30\% photoelectron heating efficiency and no extra velocity broadening. Together, these results make HD 189733b a particularly compelling test case: while it shares a similar photoelectron heating efficiency and line core absorption with HD 209458b, the observed He I line width in some observations requires invoking additional physics beyond 1D radial escape.

We note that observations of H$\alpha$ absorption in HD 189733b's atmosphere have produced conflicting results. While several studies have reported detections of excess H$\alpha$ absorption during transit \citep[e.g.,][]{Cauley_Redfield_Jensen_Barman_Endl_Cochran_2015, Cauley_Redfield_Jensen_Barman_2016}, a recent high-resolution analysis by \citet{Mounzer_Dethier_Lovis_Bourrier_Psaridi_Chakraborty_Lendl_Allart_Seidel_Osorio_etal._2025} found a H$\alpha$ absorption signal, but could not conclude it was purely planetary. The discrepancy among these studies is often attributed to contamination from stellar activity, including both intrinsic variability and rotational modulation of active regions. In particular, HD 189733 is a known active K-type star with strong chromospheric emission, which can evolve on timescales comparable to the planet’s transit duration. This makes it challenging to isolate the planetary H$\alpha$ signal from stellar effects. As a result, while H$\alpha$ remains a valuable diagnostic of the extended hydrogen atmosphere, interpreting its depth requires careful treatment of stellar contamination. In this work, we compare our model results to the reported detection from \citet{Cauley_Redfield_Jensen_Barman_2016}, which matches our model well. We note that the change in the He I transit depth from \citet{Salz_2018} to \citet{Orell-Miquel_2025} can also be due to changes in stellar activity. Previous work shows that the He I transit depth is very sensitive to changes in the stellar XUV flux \citep{Taylor2025ApJ...989...68T}. Here we matched the more recent observations by lowering the photoelectron heating efficiency from 40\% to 30\% but we could also have adjusted the stellar XUV flux to lower values consistent with a normal solar-like activity cycle.

\subsection{HD149026b} \label{sec:hd149}

Our model of HD~149026b provides an example of a clear discrepancy with the scaling relations. Given its G0\,IV host, $a{=}0.0429$~AU, and planetary size comparable to HD~209458b, the proposed scaling model \citep{Sanz-Forcada2025A&A...693A.285S} would place the He 10830 \AA\ EW for HD~149026b only a factor of $\sim$ 2.5 lower than that for HD~209458b in Figure~\ref{fig:scaling}, implying a similarly strong He I 10830~\AA\ signal. Instead, searches have yielded only non-detections with upper limits \citep{Biassoni2024A&A...682A.115B}, and our modeled transit depth also falls well below the scaling trend. \cite{Biassoni2024A&A...682A.115B} report a non-detection of He I 10830 \AA\ and place a 1-$\sigma$ upper limit on the excess transit depth of 0.34\%. Because converting this depth limit to EW$\cdot$R$_\star$ is not straightforward without assumptions about the line profile and continuum normalization, we compare our model predictions directly to the published upper limit. Here we use the same model framework as for HD~209458b and HD~189733b to diagnose the origin of this discrepency, showing that the planet’s high surface gravity compresses the thermosphere, suppresses hydrodynamic expansion, and strengthens diffusive separation—depleting helium (and hence He I 2$^3$S) at observable altitudes and naturally produces a weak or absent helium transit. Our best-fit model assumes a photoelectron heating efficiency of 35\% and yields a mass-loss rate of $\dot{M} = 1.64 \times 10^{9}$ g/s, significantly lower than the values derived for HD 209458b and HD 189733b. The temperature profile peaks at 8100 K near 1.73 $R_p$. The cyan lines of Figure~\ref{fig:all_tvcomp} shows the thermal structure and He/H ratio of the upper atmosphere. The He/H ratio decreases from 8\% at the base to less than 0.0001\% at high altitudes, pointing to significant diffusive separation, resulting in low He I (2$^3$S) densities compared to the other hot Jupiters. 




\begin{figure}[h!]
    \centering
    \includegraphics[width=0.38\textwidth]{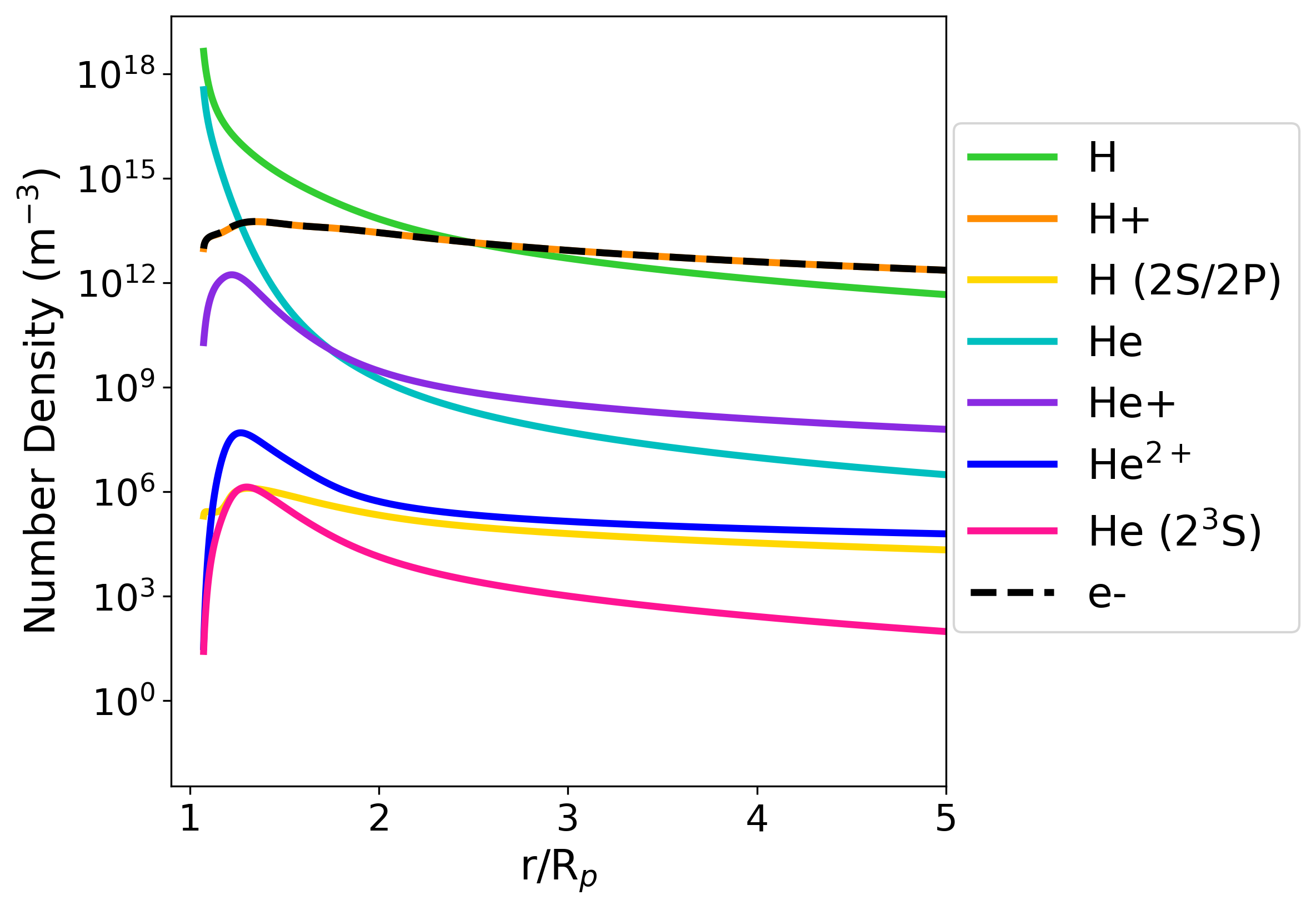}
    \includegraphics[width=0.28\textwidth]{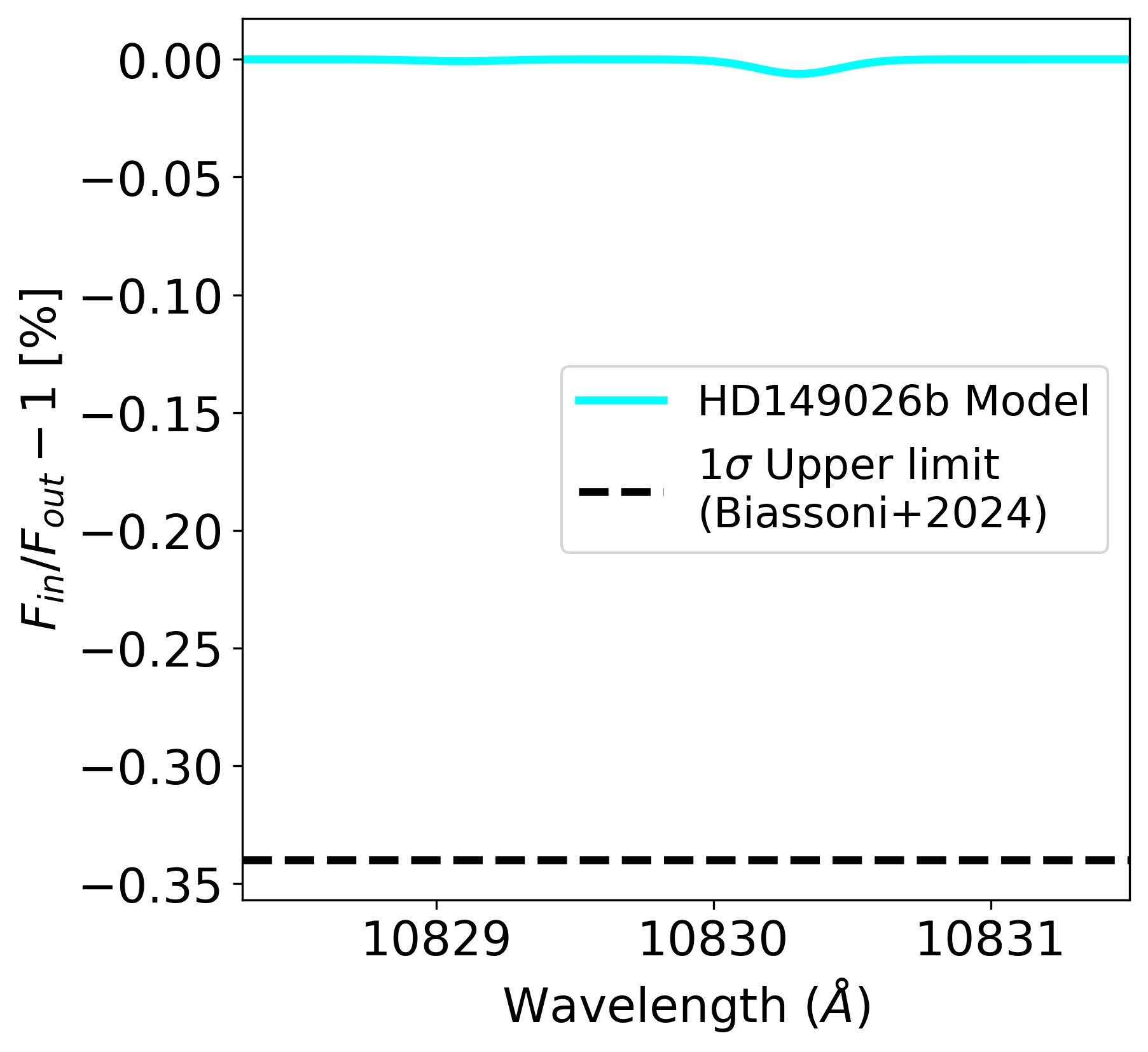}
    \includegraphics[width=0.30\textwidth]{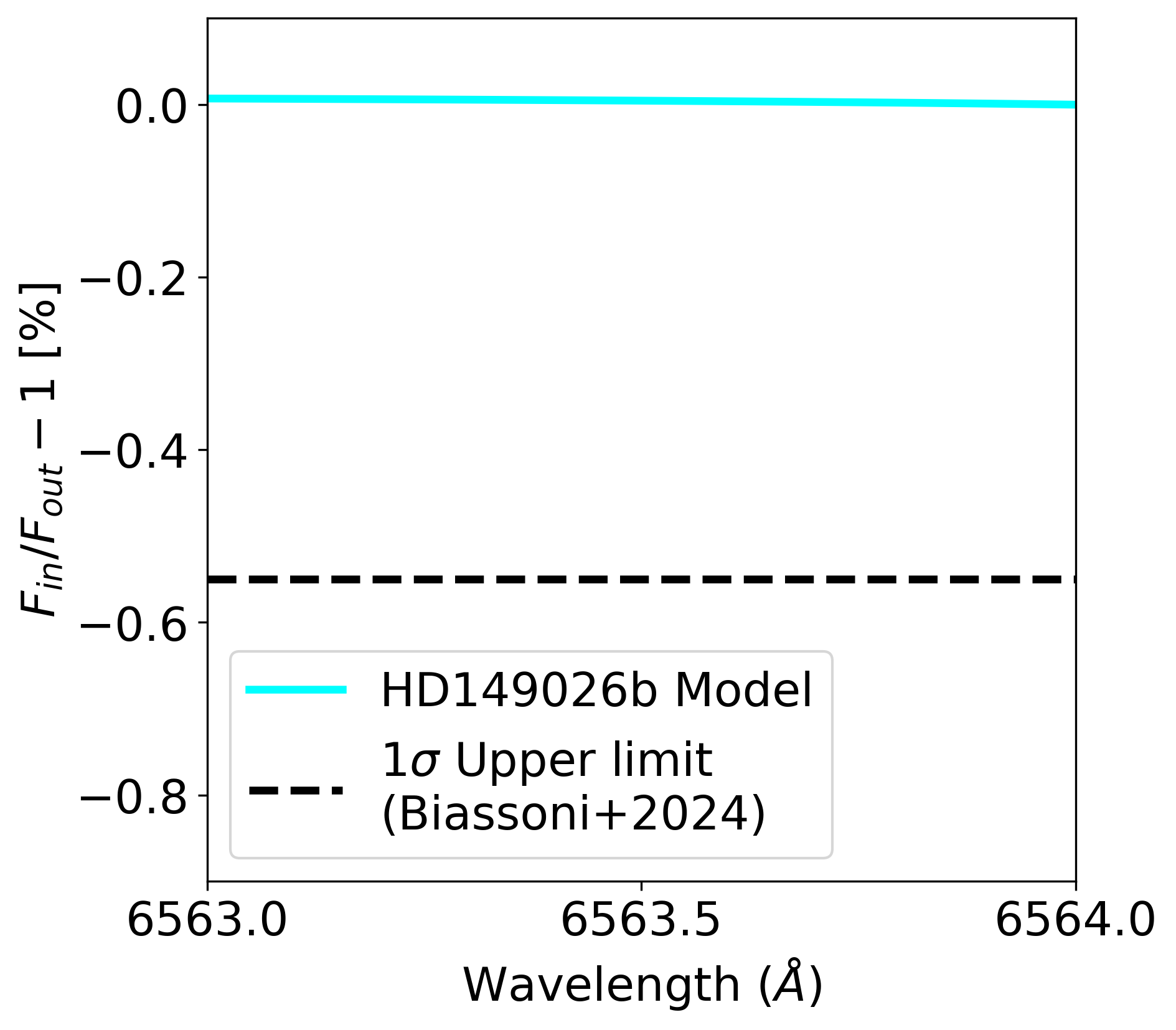}
    \caption{Full atmosphere model results for HD 149026b in the upper atmosphere. \textit{Left:} Number densities of H, H$^+$, He, He$^+$, He I (2$^3$S), and electrons. \textit{Middle:} Modeled He I transit depth spectra for HD 209458b compared with the observations from \cite{Biassoni2024A&A...682A.115B} \textit{Right:} H$\alpha$ Balmer line compared with the upper limit from \cite{Biassoni2024A&A...682A.115B}. Both features show significantly reduced transit depths compared to those in HD 209458b and HD 189733b due to the planet’s high surface gravity.}
    \label{fig:hd149_transit}
\end{figure}

As shown in Figure~\ref{fig:hd149_transit}, the modeled He I 10830 \AA\ transit depth is only $\sim$0.01\%, and we find no H$\alpha$ excess absorption, thus we predict non-detections for both lines. These values are an order of magnitude smaller than those observed or predicted for HD 209458b and HD 189733b, and for the helium transit, this suppression is directly attributable to the planet's high gravity. The steep gravitational potential well weakens escape, which strengthens diffusive separation between He/H, leading to a lower metastable population compared to the other planets. For H$\alpha$, as seen in the left panel of Figure \ref{fig:hd149_transit}, the excited hydrogen population is much less abundant at lower radii compared to HD209458b and HD189733b (see left panels of Figures \ref{fig:hd209_transit} and \ref{fig:hd189_transit}) because the incident stellar Ly$\alpha$ flux on the planet is lower than on HD 209458 b or HD189733 b. The less intense Ly$\alpha$ irradiation and weak escape thus suppress H$\alpha$ absorption. These results place our model prediction for HD 149026b in a distinct escape regime compared to typical hot Jupiters like HD209458b and HD189733b. Our model predicts that its helium and hydrogen transit depths do not conform to proposed scaling laws based on XUV flux or mass-loss rate alone, and instead reflect the strong influence of gravity in suppressing atmospheric escape signatures and enhancing diffusive separation. We note that although the surface gravity of HD 189733 b is comparable to that of HD 149026 b, the substantially weaker incident stellar XUV flux at HD 149026 b leads to a different escape/thermospheric outcome.

\subsection{GJ1214b} \label{sec:1214}

GJ 1214b is a warm sub-Neptune with a relatively low equilibrium temperature of 450 K and a high mean molecular weight atmosphere. It represents a fundamentally different regime of atmospheric escape compared to the three hot Jupiters discussed above. In this context, we use the H/He atom model as a baseline test case for a sub-Neptune irradiated by an M dwarf—consistent with many prior escape studies \citep{Lampón_2021, Orell_2022}—and then introduce a minimal molecular extension (H$_2$, H$_2^+$, H$_3^+$, HeH$^+$) to show how the presence of just H$_2$ already alters the ionization balance, cooling, and the He I 2$^3$S reservoir. This stepwise approach allows direct comparison to the hot Jupiters (identical physics, different composition) and cleanly attributes changes in the He I signal to the inclusion of H$_2$. We emphasize that GJ~1214b hosts a high–mean-molecular-weight atmosphere (e.g., H$_2$O- and/or high metallicity) \citep{Lavvas_Paraskevaidou_Arfaux_2024} with additional radiative coolants and ion–molecule pathways not included here; a fully comprehensive, high-metallicity network (including non-LTE effects, H$_2$O/CO/CO$_2$/CH$_4$ chemistry, expanded cooling, and aerosol microphysics) is beyond the scope of this paper and will be addressed in future work.

\begin{figure}[h!]
    \centering
    \includegraphics[width=0.44\textwidth]{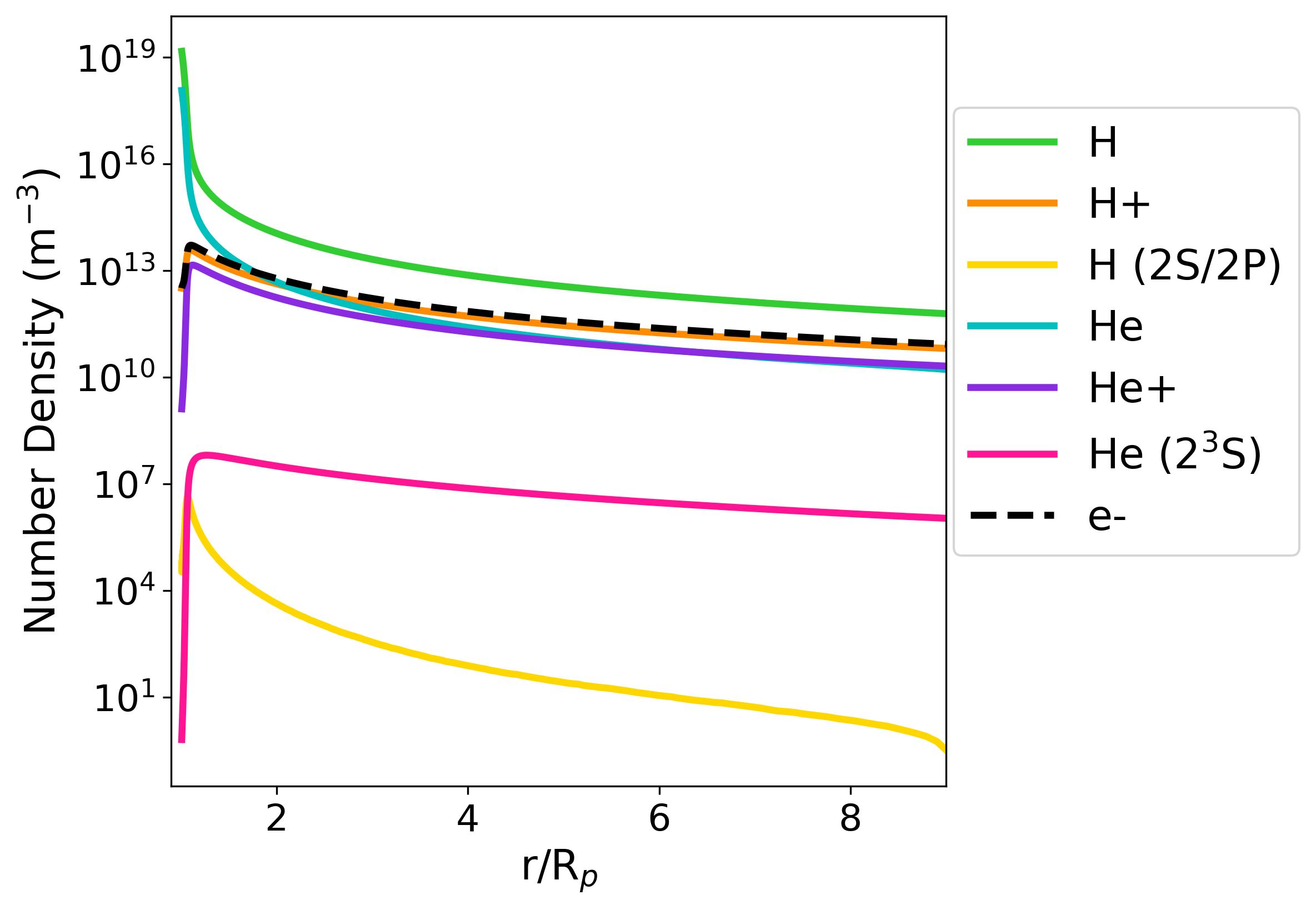}
    \includegraphics[width=0.53\textwidth]{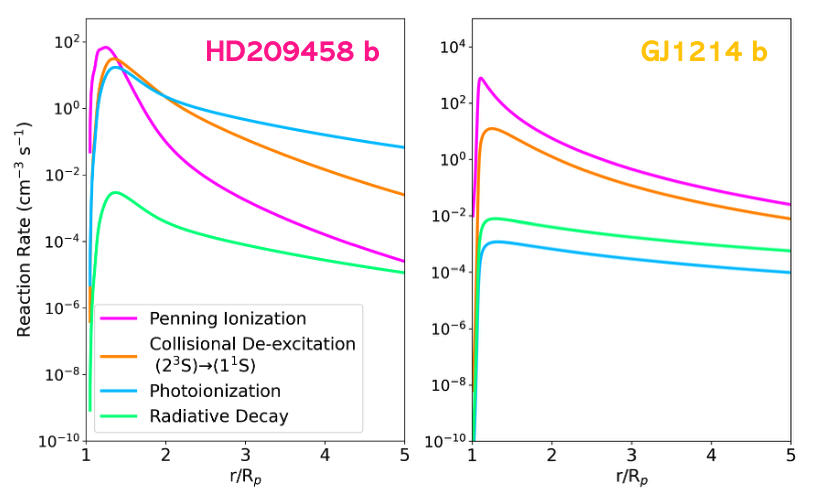}
        \caption{\textit{Left:} Atom-only model results for GJ 1214b. Species composition including H, H$^+$, He, He$^+$, He I (2$^3$S), and electrons. \textit{Right:} Comparison of metastable helium (He I 2$^3$S) loss processes in HD~209458b and GJ~1214b. For HD~209458b, the dominant loss mechanisms transition with altitude: Penning ionization dominates near the base, collisional de-excitation contributes at intermediate pressures, and photoionization becomes the primary loss pathway above $\sim$2 $R_p$. In contrast, the atmosphere of GJ~1214b remains neutral at all altitudes, such that Penning ionization with ground-state hydrogen dominates across the full vertical extent. This difference reflects the key role of ionization balance in setting the metastable helium population in hot Jupiters versus sub-Neptunes.}
    \label{fig:rates}
\end{figure}

We begin with an H/He atom-only model for consistency with previous studies. The model assumes a photoelectron heating efficiency of 35\%. The yellow lines of Figure \ref{fig:all_tvcomp} show the resulting upper atmosphere structure. The mass loss rate is $\dot{M} = 1.30 \times 10^9$ g/s, less than those of the hot Jupiters. The temperature profile peaks at 2691 K. In our hot Jupiter models, the temperatures are significantly higher.  This difference is explained by the lower gravity of the sub-Neptune that makes it more susceptible to the effects of atmospheric escape, which acts as the primary cooling mechanism above the heating peak. As shown in the left panel of Figure \ref{fig:rates}, while the thermospheres of HD 209458b and HD 189733b transition from neutral to ionized hydrogen at altitudes near 2~$R_p$, the atmosphere of GJ 1214b remains dominated by neutral hydrogen throughout the full vertical extent of the model. The prevalence of neutral H at high radii in this model is due to a combination of a relatively low temperature and high escape flux that replenishes neutral H at high altitudes where it would otherwise be predominantly dissociated and ionized, as in the hot Jupiter models.
This fundamental change in composition alters the dominant processes controlling the metastable helium population (He I 2$^3$ S).  In the hot Jupiter models, the excited helium loss is governed by a combination of Penning ionization, collisional excitation with thermal electrons, and photoionization loss, with different mechanisms dominating at different altitudes. In contrast, for GJ 1214b, where free-electron densities are comparatively low and the atmosphere remains largely neutral, the loss rate of the metastable helium population is dominated by Penning ionization with ground-state hydrogen (see right panel of  Figure \ref{fig:rates}).

Next, we include molecular hydrogen and the associated ion chemistry. The inclusion of H$_2$ alters the mean molecular weight, radiative cooling, and thermal structure of the upper atmosphere, while H$_3^+$ and HeH$^+$ affect ionization balance and energy balance. In this case, H$_3^+$ cooling becomes a significant contributor to the total radiative cooling, particularly near the temperature peak. The temperature structure is slightly different compared to the atom-only model, with a slightly higher peak temperature of 3040 K due to less efficient escape which is the dominant cooling mechanism. The mass-loss rate is 7.33 $\times 10^8$ g/s, slightly lower than the atom-only model, which is expected due to the increase in mean molecular weight. The He I (2$^3$S) abundance is lower than in the atom-only case due to a lower density of ionized helium, due to shielding from H$_2$, that helps to form the metastable state. 

\begin{figure}[h!]
    \centering
    \includegraphics[width=0.32\textwidth]{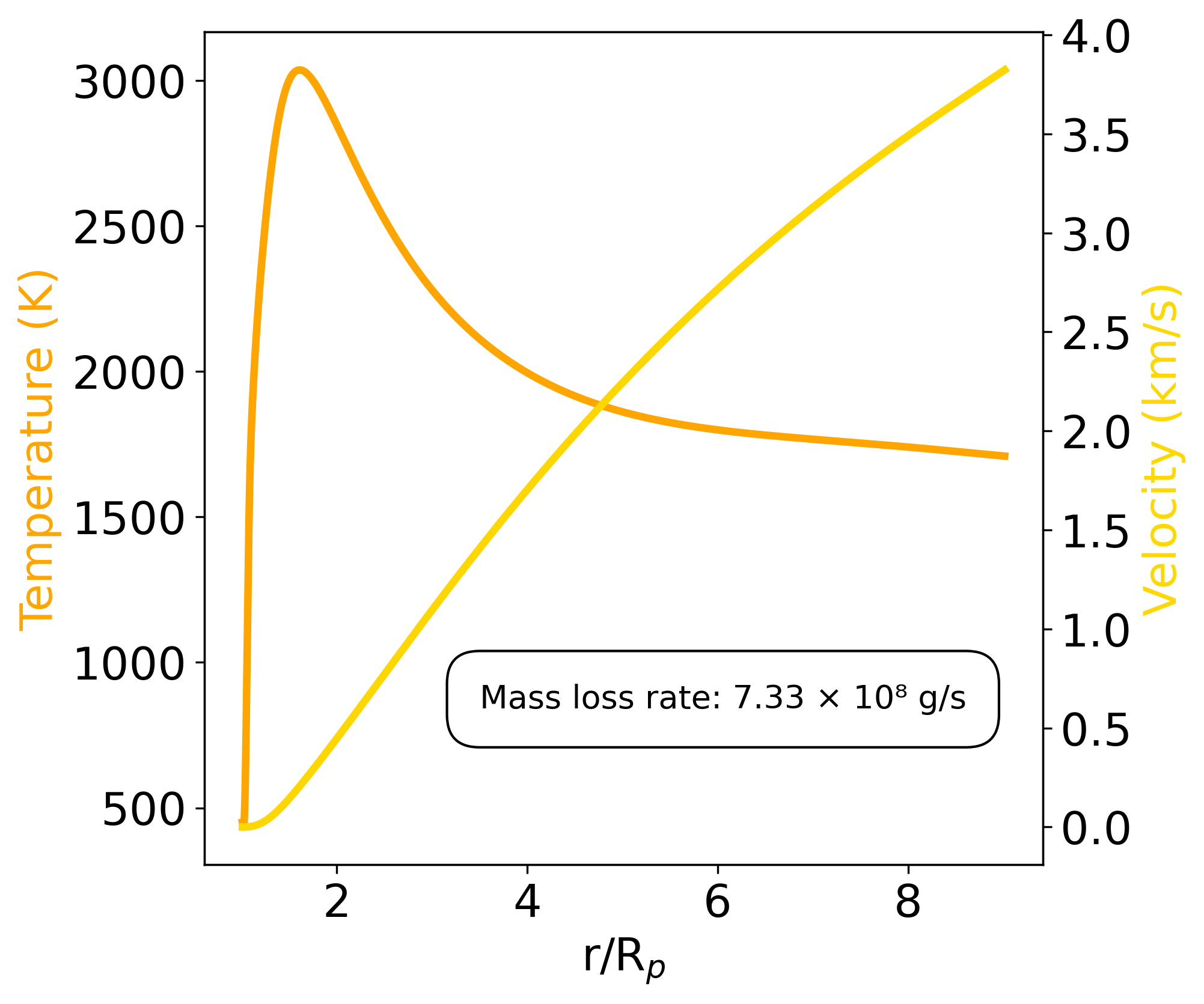}
    \includegraphics[width=0.37\textwidth]{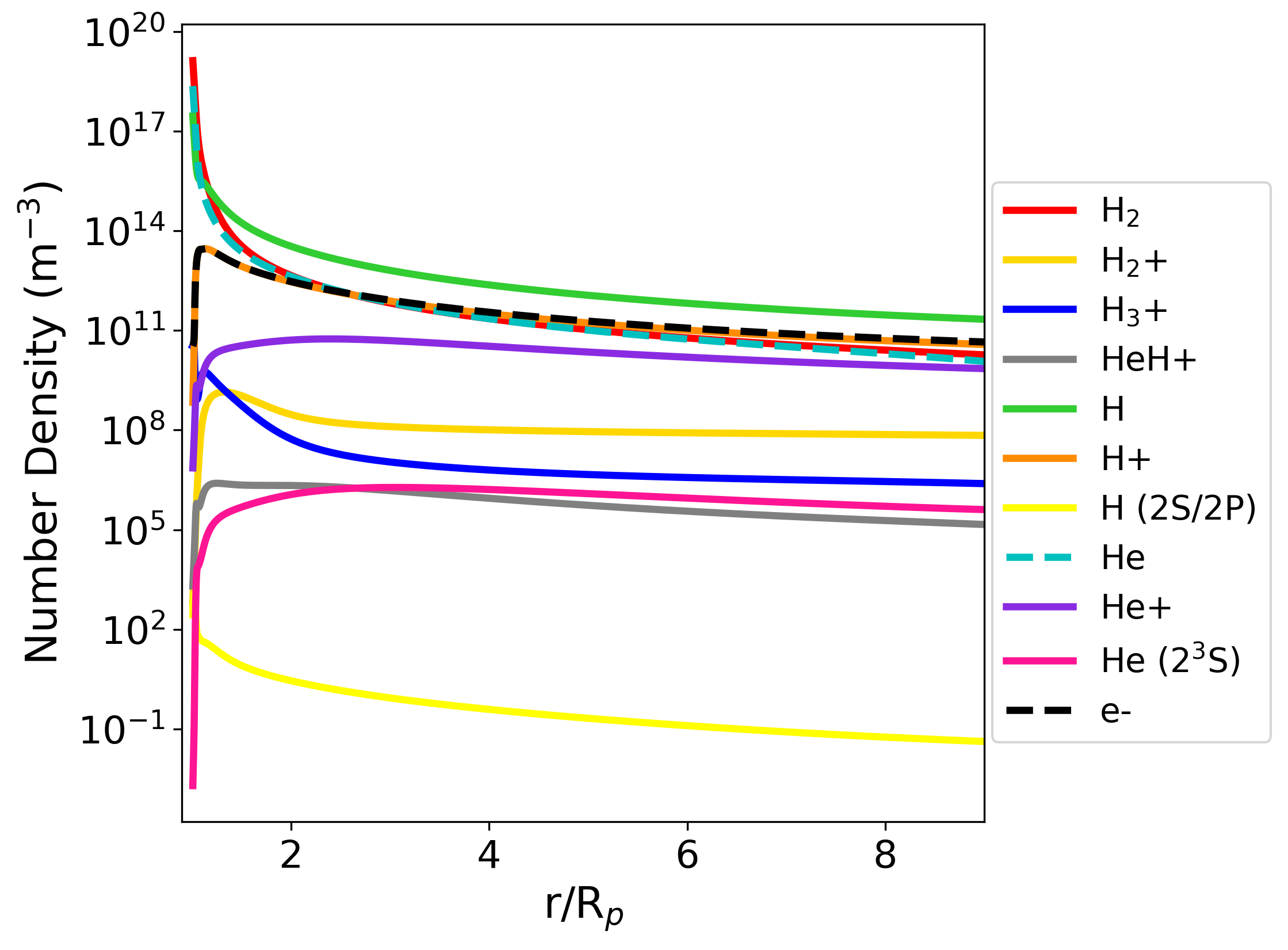}
    \includegraphics[width=0.275\textwidth]{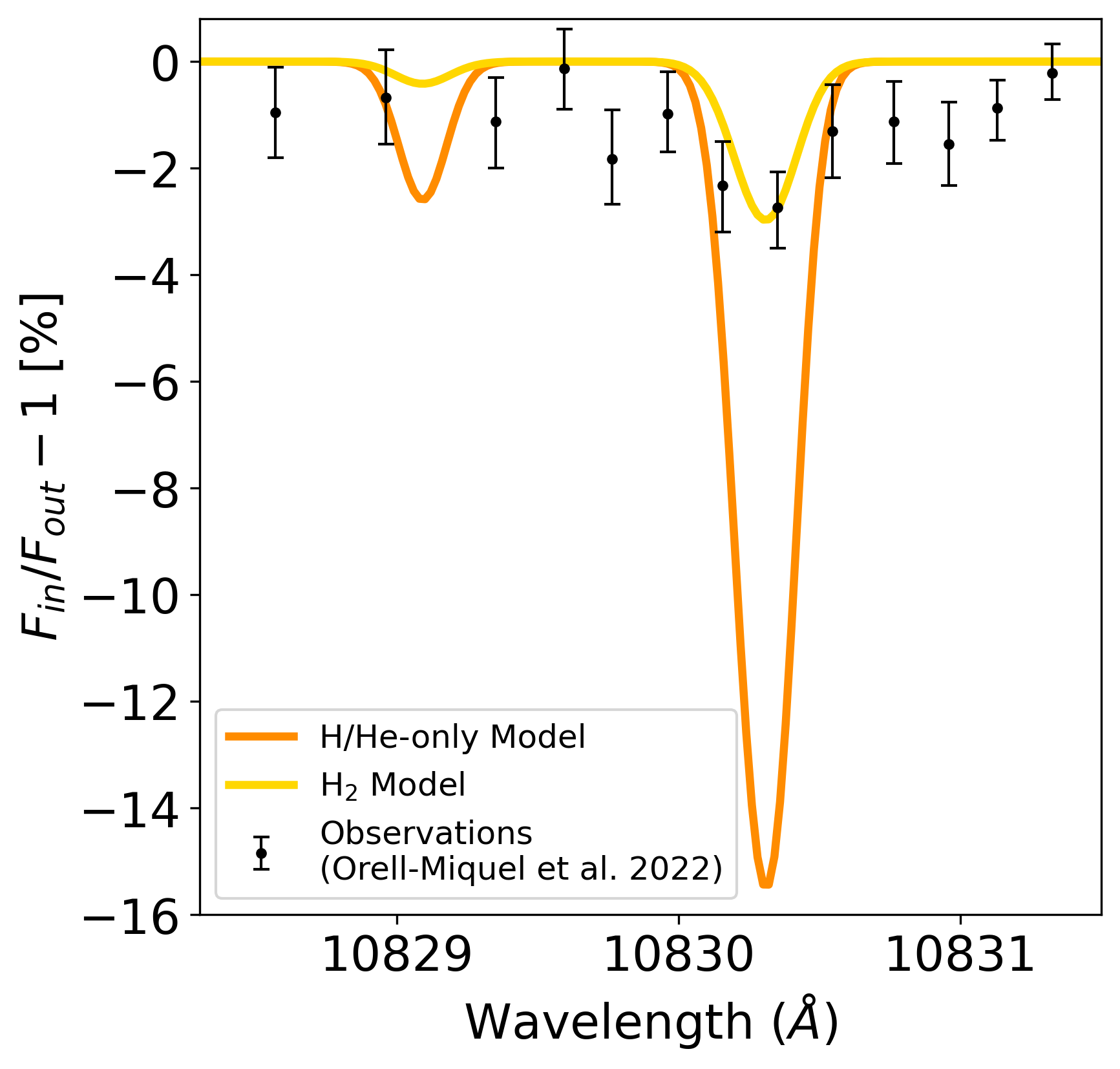}
    \caption{Molecular model results for GJ 1214b. \textit{Left:} Temperature and outflow velocity profiles. \textit{Middle:} Composition including H$_2$, H$_3^+$, HeH$^+$, and relevant atomic species. \textit{Right:} Modeled He I 10830~\AA\ transit depth spectra for GJ 1214b compared to observations from \cite{Orell_2022}. The orange curve shows the atom-only model, and the yellow curve shows the H$_2$ model. While the line shapes are similar, the inclusion of H$_2$ reduces the maximum transit depth from $\sim$14\% to $\sim$7\%. Note that we find no H$\alpha$ transit depth in this case.}
    \label{fig:gj1214_mol}
\end{figure}

The right panel of Figure~\ref{fig:gj1214_mol} shows the modeled He I 10830~\AA\ transit profiles for both compositions. The overall transit depth is larger in the atom-only model, which predicts a maximum line core absorption of $\sim$14\%, compared to $\sim$7\% in the H$_2$ model. The inclusion of molecular hydrogen reduces the metastable helium density primarily by reducing escape rates, and shielding helium from ionizing photons. In the H$_2$ model, the H and H$_2$ photoionization continua attenuate EUV photons shortward of 504\,\AA, reducing He I ground-state photoionization and thus \(n_{\rm He^+}\); ion–molecule pathways (He$^+$\,+\,H$_2$\(\rightarrow\)HeH$^+$\,+\,H; H$_2^+$\,+\,H$_2$\(\rightarrow\)H$_3^+$\,+\,H) further deplete He$^+$ and electrons. Combined with a higher-\(\mu\) thermosphere (smaller scale height), these effects lower the metastable helium column relative to the atom-only case. Consequently, even though the line shapes are similar, the H$_2$ model yields a much smaller He I signal because the available He$^+$ reservoir for populating 2$^3$S is reduced under otherwise identical irradiation conditions. We note that published He I 10830 \AA\ measurements of GJ 1214 b are not consistent across epochs, with one reported detection alongside multiple non-detections and upper limits, suggesting that the apparent He I transit depth may be variable \citep{Kasper_2022, Orell_2022, Allart_2023, masson2024probing}. Given this observational heterogeneity, we interpret our predictions as representative of a steady-state/typical atmosphere under the assumed irradiation, and we compare primarily to the tentative detection of He I 10830 \AA\ given in \cite{Orell_2022}. Note that we find no H$\alpha$ transit depth in this case, or in the He/H only model. The stellar Ly$\alpha$ profile incident on this planet is much weaker than for the hot-Jupiters, yielding Ly$\alpha$ mean intensities that are uniformly low throughout the thermosphere. Combined with the cool, largely neutral upper atmosphere, this keeps the excited state hydrogen population negligible and prevents an observable H$\alpha$ transit depth.


\section{Discussion} \label{sec:dis}

\noindent In this work, we present simulations of the escaping atmospheres of three well-known hot Jupiters orbiting G and K-type host stars and one well-known sub-Neptune orbiting an M-type host star to investigate how the observed He I 10830~\AA\ and H$\alpha$ transit depths depend on host star properties and system parameters. Our goal was to reach general conclusions about what processes explain the scatter and outliers particularly in the existing He I 10830~\AA\ transit observations. Empirical scaling relations used to predict the He I 10830~\AA\ transit depth typically tie the signal to energy-limited escape via stellar XUV forcing—e.g., trends with $F_{\rm XUV}$ alone or with $F_{\rm XUV}/\rho_p$ motivated by simple mass loss arguments—and have been used in recent uniform reanalyses of the helium observations \citep[e.g.,][]{McCreery_DosSantos_Espinoza_Allart_Kirk_2025,  Krishnamurthy_2024}. At the same time, synthetic population studies show that He(2$^3$S) absorption correlates only weakly with $F_{\rm XUV}$ and depends sensitively on the stellar SED shape and system geometry, motivating multi-parameter prescriptions that incorporate $F_{\rm XUV}$, $R_p$, $M_p$, $R_\star$, and the balance of He-ionizing to He$^\ast$-pumping photons \citep[e.g.,][]{Linssen2024A&A...688A..43L, Krishnamurthy_2024,Sanz-Forcada2025A&A...693A.285S}. 

Consistent with this background and the most recent scaling relations, we compare our detailed model results with the scaling relation of \citet{Sanz-Forcada2025A&A...693A.285S}, given by 
\begin{equation}
\mathrm{EW}\,R_\star^2 \;\propto\; \frac{F_{\rm XUV,He}\,R_p^{2}}{\Phi_p}.
\label{eq:sanz}
\end{equation}
where $F_{\mathrm{XUV,He}}\equiv F(1$–$504\,\text{\AA})$, which governs He I ground-state photoionization and thus the recombination pathway to the He (2$^3$S) state, $\Phi_p \equiv GM_p/R_p$, and $R_*$ is the host star radius. We note that the right-hand side of this scaling law is closely connected to energy-limited mass loss rate, often estimated as
\begin{equation}
\dot{M}_{\rm lim} \;=\; \frac{\epsilon\,\pi R_{p}^{2}\,F_{\rm XUV,H}}{\Phi_p},
\end{equation}
where $F_{\mathrm{XUV,H}}\equiv F(1$–$912\,\text{\AA})$ governs photoionization of H. For all the spectral types (G/K/M) included in our study, most of the stellar XUV flux is at $\lambda <$~504~\AA. Also, the energy from photoionization is transferred to the atmosphere by photoelectrons with energies that increase with decreasing photon wavelength. We estimate the heating impact of different wavelength ranges by first subtracting 13.6 eV, the ionization potential of H, from each photon energy and then integrating the energy flux in our spectra. We find that the ratio of the energy available to heat the atmosphere in the 1--504~\AA~range to the 504--912~\AA~range is $\sim$11 for HD209458b and HD189733b, $\sim$22 for HD149026b, and $\sim$496 for GJ1214b. Thus, the scaling law in equation~(\ref{eq:sanz}) can be written as
\begin{equation}
\mathrm{EW}\,R_\star^2 \;\propto\; \frac{\dot{M}_{\rm lim}}{\epsilon\,\pi},
\end{equation}
which provides a useful additional shorthand for understanding our results and a connection to scaling laws based on the mass-loss rate.

In Figure~\ref{fig:scaling}, we compare our modeled equivalent widths to the scaling relation in \cite{Sanz-Forcada2025A&A...693A.285S}, our Equation 2. The results for HD~209458b and GJ~1214b lie close to the predicted scaling relation, consistent with expectations from their irradiation levels and observed line depths. However, we caution against overinterpreting the apparent agreement for HD~209458b and GJ~1214b. For GJ~1214b, published He I observations are inconsistent—with both detections and non-detections reported—so its “match” to the scaling relation is sensitive to which dataset is adopted \citep[e.g.,][]{Kasper_2022, Orell_2022, Allart_2023, masson2024probing}. Moreover, our GJ 1214 b models are highly sensitive to the assumed atmospheric composition, yielding substantially different He I absorption predictions across plausible compositions; thus, our inferred placement relative to the scaling relation should be interpreted with caution. For HD~209458b, its status as the archetypal hot Jupiter means many empirical recipes may be benchmarked or tuned against it, so agreement there may partly reflect construction. In short, even where the scaling “works,” the concordance is not without caveats—reinforcing the need to interpret observations with physics-based models. The transit depth for HD~189733b falls below the scaling relation. The enhanced XUV flux from its active K-type host is expected to yield a deeper He I 10830~\AA\ absorption signal than for HD~209458b. The absorption line profile for HD189733b in many observations is significantly broader than for HD209458b, but the core transit depth is similar for both planets. In the most recently published observations of HD189733b, both the transit depth and line profile are similar to HD209458b \citep{Orell-Miquel_2025}. For HD149026b, our models predict similarly weak He I absorption, placing it well below the scaling relation as well. Our predicted transit depth for HD~149026b show that the high surface gravity suppresses atmospheric expansion and enhances diffusive separation of hydrogen and helium, producing a weak helium signal. These findings demonstrate that compact or gravitationally bound thermospheres can significantly limit He~I absorption even under strong irradiation. 
Our self-consistent model reproduces the observed He~I and H$\alpha$ depths and broadly satisfies the upper limits for the hot Jupiters because key processes—thermal structure, ionization balance, diffusion, and recombination—are explicitly included. This highlights that scaling relationships based solely on XUV flux and planetary and stellar radius cannot explain the scatter in the observed He I transit depths at the population level, and that physically detailed models are still essential to interpreting helium transit depths in individual systems. Beyond reproducing the measurements, physically detailed models provide direct constraints on the underlying atmospheric physics—e.g., the temperature structure, ionization state, and the relative importance of transport versus chemistry—that set the metastable helium population. In contrast, more parametric or isothermal frameworks can often fit a given depth by adjusting effective temperatures or mass-loss rates as free parameters, but those values are not necessarily unique or physically self-consistent. Our approach, therefore, links the observed transit depths to a realistic thermospheric solution, rather than to tunable parameters.

Given the approximate dependency of the scaling relation on the energy-limited escape rate, a comparison of our model results for the hot Jupiters and the energy limit is also warranted. For HD209458b, our best-fit atoms only model from \citet{Taylor2025ApJ...989...68T} provides an excellent fit to different observations \citep{Alonso-Floriano2019,Orell-Miquel_2025}. This model uses a photoelectron heating efficiency of 40\% and yields a mass-loss rate of $\dot{M} = 1.9 \times 10^{10}$ g/s. The net heating efficiency of the model is 15\%, which is the fraction of the star’s EUV energy that actually becomes heat in the planet’s upper atmosphere after immediate radiative and other losses are removed. Our model's mass-loss rate is roughly consistent with the energy limit provided that we replace $R_p$ in the formula with 1.32 $R_p$, which lies between the heating peak (1.07 $R_p$) and temperature peak (1.7 $R_p$) in our model. The addition of heavy elements to the model with solar abundances did not significantly affect the temperature and velocity structure. It did, however, change the He I 10830~\AA\ transit depth and required us to lower the best-fit photoelectron heating efficiency to 23\%, yielding a similar mass loss rate as the H/He only model \citep[see][for further details]{Taylor2025ApJ...989...68T}. This shows that the presence of heavy elements with solar abundances is not critically important to models of atmospheric escape from similar hot Jupiters, perhaps fortunately, given the complexity of the models that include them.

For HD189733b, the incident stellar XUV flux at 0-91.1 nm is 18 times higher than on HD209458b. In this case, we obtain a mass loss rate of 7.1~$\times$~10$^9$~g~s$^{-1}$ and a net heating efficiency of 0.03 \%. This yields an energy-limited mass loss rate of 8.8~$\times$~10$^{9}$ g~s$^{-1}$, roughly similar to our model-derived rate. Note that the 0.03/0.15 difference in the net heating efficiency between HD209458b and HD189733b shows that the energy-limited escape rate can vary significantly. Our best-fit model, however, uses a photoelectron heating efficiency of 40\% that is similar to HD209458b. The range of 20--40\% is roughly consistent with solar system aeronomy. Peak midday heating efficiencies inferred for Earth’s thermosphere are $\sim$50–55\% \citep{Torr1980JGR, Richards2012CJP}, and gas-giant thermosphere studies commonly adopt comparable values when modeling energy deposition and global circulation \citep[e.g.,][]{Strobel2012Icarus,Shematovich2014AA}. For HD~189733b, we included H$_2$ for consistency with the lower/middle atmosphere, but the strong XUV field and hot thermosphere dissociate H$_2$ at very low altitudes—well below the line‐forming region—so molecular shielding and H$_3^+$ cooling are negligible and the addition of H$_2$ has only a small impact on the mass-loss rate or on the modeled He I and H$\alpha$ transit depths. Diffusive separation of He on HD189733b is more efficient at the lowest altitudes in the thermosphere while the He/H ratio at higher altitudes is similar to HD209458b. This is in line with the factor of two difference in mass loss rates, both of which exceed the cross-over mass limit \citep{Taylor2025ApJ...989...68T}. Matching the more recent observations from \citet{Orell-Miquel_2025} requires either a lower photoelectron heating efficiency or incident stellar XUV flux. We chose to fit the observations by lowering the photoelectron heating efficiency to 30\% as a proof of concept.  The He I 10830~\AA\ transit depth, however, changes by a factor of about five based on changes to XUV fluxes following the solar activity cycle for HD209458 \citep{Taylor2025ApJ...989...68T}. HD189733 is a much more active star than the sun and the more recent observations could also be explained by lowering the stellar XUV flux. We conclude that the main reasons for why the He I 10830~\AA\ transit depth does not match with the scaling relations are the planet's higher gravity and incident XUV flux compared to HD209458b.

HD~149026b presents an especially interesting case in the context of proposed helium scaling relations. As a hot Jupiter orbiting a G0 subgiant with similar irradiation to HD~209458b, scaling relationships such as those in \citet{Sanz-Forcada2025A&A...693A.285S} predict that it should exhibit a stronger He~I 10830~\AA\ absorption signal that our model predicts. However, observations have failed to detect the line, and the 1-$\sigma$ upper limit on the transit depth is relatively low \citep{Biassoni2024A&A...682A.115B}, suggesting that additional physics not captured by the scaling relationships dominate in this system. The incident XUV flux on HD149026b is 1.0 W~m$^{-2}$ i.e., lower than on HD209458b by a factor of 1.6. The photoelectron heating efficiency is effectively unconstrained because the observations provide only upper limits on the He I 10830~\AA\ transit. We use 35\% as an average of the values adopted for HD209458b and HD189733b. The mass loss rate predicted by our model is 1.64~$\times$~10$^{9}$ g~s$^{-1}$. The expected energy-limited mass loss rate for a net heating efficiency of 11\% is 1.2~$\times$10$^9$ g~s$^{-1}$, roughly consistent with our model. This mass-loss rate is 12 times lower than the mass loss rate for HD209458b and below the cross-over mass limit for He. Our model predicts that diffusive separation of He in the atmosphere of HD149026b is very strong and explains the non-detection of the He I 10830~\AA\ transit, and the planet's departure from the scaling relation. The key difference to the other hot Jupiters in our study is the relatively high surface gravity of HD149026 b ($g \sim 23$ m s$^{-2}$) that acts as a break on hydrodynamic escape. This suppressed escape enhances the role of molecular diffusion. With lower bulk outflow, diffusive separation of hydrogen and helium becomes efficient over a larger vertical range. Consequently, the modeled He~I transit depth is only $\sim$0.01\% and no significant H$\alpha$ absorption is predicted. This follows from the planet’s high gravity, which produces a compact, weakly escaping thermosphere where (i) diffusive separation depletes He at line‐forming altitudes and lowers $n_e$ and He$^+$, suppressing recombination into the metastable 2$^3$S state, and (ii) the reduced scale height and Ly$\alpha$ mean intensity keep the H($n{=}2$) population small and rapidly quenched—leaving too little column above the limb for detectable absorption in either line. To confirm this explanation, we run a model of HD189733b, which has similar surface gravity to that of HD149026b (22 m s$^{-2}$), under the same irradiation condition as HD149026b. This test confirms that when gravity is relatively high and irradiation is relatively low (the incident XUV flux is $\sim$30 times lower than on HD189733b), the upper atmosphere escapes less efficiently, diffusive separation increases, and the metastable helium population significantly decreases. Additionally, repeated He I 10830~\AA\ transit observations of HD~149026b would be particularly valuable to robustly establish whether the line is truly absent or present at a low level, thereby providing a decisive test of our predicted weak absorption for this system.

The He I 10830~\AA\ absorption line profile in many observations of HD189733b is much broader than the line profile on HD209458b \citep{Salz_2018,Zhang_Knutson_Wang_Dai_Barragan_2022,Allart_2023,masson2024probing}. Our model does predict a line that is intrinsically broader than for HD~209458b, owing to stronger thermal and rotational broadening, but this additional broadening remains insufficient, underestimating the observed line width by a factor of $\sim$2--3. We tested variations in the free parameters (heating efficiency, $K_{zz}$, stellar activity level/XUV input, base composition, and lower-boundary placement): configurations that increase the width either deepen the core transit depth too much or violate the H$\alpha$ constraints, and none simultaneously reproduce both core depth and width. Processes outside of our 1D radial framework, such as day–night or zonal winds at the terminator, are a natural next consideration, but typical GCM wind speeds (few km\,s$^{-1}$) are unlikely to supply the additional $\sim$12~km\,s$^{-1}$ required to match the observed width. \textcolor{magenta}  Furthermore, the presence of fast horizontal winds is predicted for both HD209458b and HD189733b in the middle atmosphere ($\sim$ millibar), and the differences in predicted wind speeds do not provide a solution. {Other 1D models that have attempted to reproduce the He I 10830~\AA\ absorption line profile have also had to impose additional turbulent broadening to match the width of the profile \citep{Lampón_2021}.} Previous work by \citet{Rumenskikh_Shaikhislamov_Khodachenko_Lammer_Miroshnichenko_Berezutsky_Fossati_2022}, who used a 3D multi-fluid H/He model, also explored this problem. They imposed an empirical reduction to H line cooling in order to increase the mass loss rate and broaden the line profile. They also imposed faster zonal (equatorial) jets by prescribing additional rotation, matching the observed width with equatorial velocities of order 10--20 km~s$^{-1}$, similar to our required additional velocity broadening. Finally, they required a very low He/H ratio of 0.005 in the upper atmosphere to match the observations. Such a low He/H ratio is not supported by any planet formation theories and cannot represent the deep helium abundance of the planet. Additionally, our model that includes diffusive separation of helium self-consistently does not support a ratio this low either. Instead, our model comfortably matches the core transit depth and we consider additional velocity broadening as a better option for matching the observed line width than the combination of the options from \citet{Rumenskikh_Shaikhislamov_Khodachenko_Lammer_Miroshnichenko_Berezutsky_Fossati_2022}. Our model has no trouble matching both the line core and width from the more recent observations of \citet{Orell-Miquel_2025}, so this point of view is well justified.

At the heart of this problem is the similarity of the two planets, HD209458b and HD189733b. Although there are differences between the incident XUV flux, the host star, the orbit, and planet properties, all of these differences are included in our model and yet we cannot always reproduce the line profile for HD189733b. Multi-dimensional models have also explored these systems and cannot self-consistently reproduce the observations \citep{Rumenskikh_Shaikhislamov_Khodachenko_Lammer_Miroshnichenko_Berezutsky_Fossati_2022}. One aspect of the planets may, however, be different. The magnetic field of HD189733b is expected to be much stronger than on HD209458b. Observations and models of the upper atmosphere of HD209458b have been used to propose an upper limit of about 0.1 G on the equatorial magnetic field strength of the planets \citep{Kislyakova_Holmström_Lammer_Odert_Khodachenko_2014,Khodachenko2021MNRAS.507.3626K}. For HD~189733b, star–planet interaction modeling of Ca\,\textsc{ii}~K variability implies a surface polar dipole of $B_{\rm p}=20\pm7$~G, with energy-flux dynamo scalings yielding $\sim$53$\pm$17~G for comparison \citep{Cauley_Shkolnik_Llama_Lanza_2019}. We have introduced the additional broadening to match the observed He I 10830~\AA\ line profile essentially as turbulent, Gaussian broadening. Given that excited He is mostly produced by recombination of helium ions, this broadening can be driven by plasma turbulence. A particularly interesting idea is that HD189733b hosts a population of trapped ions on low-latitude magnetic field lines that form an energetic, turbulent plasma population around the planet. During periods of higher stellar activity, the turbulence is strengthened, similarly to the Earth's magnetosphere during high solar activity. During quiescent stellar activity, the line profile returns to "normal" and our model is able to explain it without additional turbulent broadening. While these ideas are speculative, we encourage further exploration of plasma effects and magnetic field interactions to characterize He I 10830~\AA\ transit depths and light curves.

For the warm sub-Neptune GJ~1214b, our results are fundamentally different from those for the hot Jupiters, even for models that only include H and He. The upper atmosphere of GJ~1214b remains dominated by neutral hydrogen at all altitudes and Penning ionization dominates the loss rate of the metastable helium population everywhere. The prevalence of neutral H agrees with previous simulations of Neptune-type planets, whether around M-type or sun-like stars \citep{parkeloyd17,2022ApJ...929...52K}. Including molecular hydrogen and the associated chemistry of H$_3^+$ and HeH$^+$ molecular ions further alters the ionization balance and suppresses the metastable helium density by shielding helium from photoionization. These effects significantly reduce the predicted He~I transit depths. We limit our models to H/He chemistry in this work for comparison with the hot Jupiters, the context from previous models, and because the effect of the different planetary properties and host star irradiation was sufficiently interesting for stand-alone results. For GJ~1214b specifically, however, transmission-spectrum fits indicate a high-metallicity atmosphere (likely tens to thousands of times solar), consistent with a water- and carbon–nitrogen–bearing composition \citep[e.g.,][]{Lavvas_Paraskevaidou_Arfaux_2024}. This implies that molecules such as H$_2$O, CO$_2$, CH$_4$, NH$_3$, and N$_2$ are abundant at (and above) our 1~$\mu$bar lower boundary, where they can (i) raise the mean molecular weight, (ii) add powerful IR cooling channels (iii) increase FUV/NUV opacity that shields He from ionization, and (iv) increase photoelectron populations —each of which directly affects the He I 10830~\AA\ signal. The related chemistry effectively links He I observations on GJ~1214b to the evidence for muted/flat transmission spectra consistent with high mean–molecular–weight atmospheres and/or aerosols \citep{Berta2012, Kempton_Zhang_Bean_Steinrueck_Piette_Parmentier_Malsky_Roman_Rauscher_Gao_etal._2023, Ohno2025ApJ...979L...7O}. We expect that increasing the heavy-molecule content lowers the He I signal by reducing He$^+$ production and the recombination pathway to the He~(2$^3$S) state, even under identical irradiation. This leads to a simple, testable hypothesis: planets with high-$\mu$ (metal-enriched or water-rich) compositions should exhibit systematically weaker He I~10830\,\AA\ absorption than H/He-dominated atmospheres at comparable $F_{\rm XUV}$ and gravity. Conversely, a strong He I detection on a sub-Neptune would imply either lower $\mu$ or additional ionization pathways that boost He$^+$ despite high-$\mu$ conditions. Coordinated campaigns that pair He I transits with contemporaneous JWST retrievals of terminator composition can therefore discriminate between high-$\mu$ and H/He–dominated regimes and quantify how composition modulates escape diagnostics.

Observations of H$\alpha$ transit depths offer an additional probe of the upper atmosphere and coordinated observations would help to resolve many degeneracies in interpreting the He I 10830~\AA\ transit signatures alone, as demonstrated by \citet{Taylor2025ApJ...989...68T}. The interpretation of the observations, as usual with upper atmosphere probes, however, is complicated due to non-LTE considerations and the observations themselves are also confusing. For both HD209458b and HD189733b, the observations are inconclusive on what the transit depth is and even whether the transit is detected or not. Given that the H$\alpha$ transit depths are much less sensitive to many of the free parameters in our model than the He I 10830~\AA\ transit depths \citep{Taylor2025ApJ...989...68T}, the instability of the transit depths is likely due to the direct effect of stellar variability on the transit depths inferred from the observations rather than the impact of the same variability on the planetary atmosphere and predicted absorption. With that caveat, our results for HD209458b are consistent with some upper limits on the H$\alpha$ transit depth \citep[e.g.,][]{jenesen2012ApJ...751...86J} while we exceed other limits \citep{Astudillo-Defru2013A&A...557A..56A,Casasayas-Barris2021A&A...647A..26C}. Our best-fit models are in excellent agreement with the H$\alpha$ observations of HD189733b \citep{Cauley_Redfield_Jensen_Barman_2016}. We predict no detectable transit at H$\alpha$ for HD149026b and GJ1214b, which further complicates the picture for this diagnostic. In both cases, the likely explanation is the slightly lower incident stellar XUV and Lyman~$\alpha$ fluxes (compared to HD209458b), combined with an insufficient temperature in the upper atmosphere to support a significant thermal population of excited H. This likely means that H$\alpha$ transits are only usable as a diagnostic for ultra-hot and hot Jupiters and even in these cases, their connection to mass loss rates and other properties of the upper atmosphere is tenuous. This further supports the view that simultaneous constraints from multiple absorption and if possible, emission lines, are the best way to really study the upper atmospheres of exoplanets.

\section{Conclusions}
\label{sec:conclusions}

Our comparative, full-atmosphere models of HD~209458b, HD~189733b, HD~149026b, and GJ~1214b place recent helium-scaling prescriptions in a broader physical context. Empirical relations between He I~10830~\AA\ strength and combinations of $F_{\rm XUV}$, planet size, and gravity appear to capture some collective behavior across samples, but individual systems frequently depart from those trends for well-defined physical reasons. Our models account for these departures by resolving thermal balance, non-LTE populations, diffusion, and composition. For the benchmark HD~209458b, we reproduce the observed He I and H$\alpha$ absoprtion without ad-hoc composition assumptions, providing a useful reference point. HD~189733b exhibits line-core depths comparable to HD~209458b, yet markedly broader He I profiles in many datasets; matching those widths requires $\sim$12~km\,s$^{-1}$ of additional non-thermal broadening, potentially consistent with magnetically mediated dynamics or turbulence. HD~149026b may illustrate a different departure: we predict that its high surface gravity suppresses hydrodynamic expansion and enhances diffusive separation, naturally yielding very weak He I despite comparable irradiation. This prediction should be tested by repeated observations that definitively detect the He I~10830~\AA\ line or confirm a non-detection for this planet. For the warm sub-Neptune GJ~1214b, H/He-only models overestimate the helium signal; introducing H$_2$ and associated ion chemistry reduces the metastable helium reservoir by altering ionization balance and cooling, producing shallower He I and no H$\alpha$. Notably, we find that each planet's upper atmosphere shows diffusive separation between hydrogen and helium, which could explain why previous studies have had to assume sub-solar H/He ratios in order to match helium transits. Taken together, these results suggest that while scaling relations can be useful for population-level intuition, reliable interpretation of individual systems—particularly outliers—benefits from physics-rich models that resolve energy balance, multi-species transport (including diffusion), and composition-dependent chemistry.

\vspace{0.5cm}

\noindent A.R.T. and T.T.K acknowledge support by NASA/XRP 80NSSC23K0260. C.H. is sponsored by Shanghai Pujiang Program (grant NO. 23PJ1414900). P.L. acknowledges support from the Programme National de Planétologie of CNRS/INSU under the project ``Temperate Exoplanets''.

\software{Astropy \citep{2018AJ....156..123A, 2022ApJ...935..167A}, Matplotlib \citep{Hunter2007}, NumPy and SciPy \citep{2020NatMe..17..261V}, Pandas \citep{mckinney-proc-scipy-2010}}

\bibliography{refs.bib}{}
\bibliographystyle{aasjournal}

\end{document}